\documentclass[acmsmall]{acmart}
\usepackage{siunitx}
\usepackage[linesnumbered,ruled,lined]{algorithm2e}
\usepackage{booktabs}
\usepackage{subcaption}
\usepackage{multirow}
\usepackage{amsmath}
\usepackage{graphicx}
\usepackage{pifont}
\usepackage{array}
\usepackage{makecell}
\usepackage{setspace}
\usepackage{caption}
\usepackage{url}
\usepackage{enumitem}
\usepackage{tabularx}
\usepackage{microtype}
\usepackage{tcolorbox}
\usepackage{threeparttable}
\usepackage[table]{xcolor}
\usepackage{hyperref} 
\usepackage{cleveref} 

\hypersetup{colorlinks = true, 
	    linkcolor = blue, 
	    urlcolor = blue,
            citecolor = red} 
\restylefloat{figure}
\AtBeginDocument{%
  }

\setcopyright{cc}
\setcctype{by}
\acmDOI{10.1145/3797083}
\acmYear{2026}
\acmJournal{PACMSE}
\acmVolume{3}
\acmNumber{FSE}
\acmArticle{FSE055}
\acmMonth{7}
\received{2025-09-12}
\received[accepted]{2025-12-22}
\begin{document}

\title{Neuron-Guided Interpretation of Code LLMs: Where, Why, and How?}

\author{Zhe Yin}
\orcid{0009-0005-1589-6254}
\affiliation{%
  \institution{Shanghai Jiao Tong University}
  \city{Shanghai}
  \country{China}
}
\email{yin_zhe@sjtu.edu.cn}

\author{Xiaodong Gu}
\orcid{0000-0002-0529-6408}
\affiliation{%
  \institution{Shanghai Jiao Tong University}
  \city{Shanghai}
  \country{China}
}
\email{xiaodong.gu@sjtu.edu.cn}

\author{Beijun Shen}
\authornote{Corresponding author.}
\orcid{0000-0001-8370-3956}
\affiliation{%
  \institution{Shanghai Jiao Tong University}
  \city{Shanghai}
  \country{China}
}
\email{bjshen@sjtu.edu.cn}
\renewcommand{\shortauthors}{Zhe Yin et al.}
\newcommand{\ourmethod}[0]{\textsc{XYZ}\xspace}
\newcommand{\todoc}[2]{{\textcolor{#1}{\textbf{[#2]}}}}
\newcommand{\todoblue}[1]{\todoc{blue}{\textbf{#1}}}
\newcommand{\todored}[1]{\todoc{red}{\textbf{#1}}}
\newcommand{\yin}[1]{\todoblue{yin: #1}}
\newcommand{\gu}[1]{\todored{gu: #1}}
\newcommand{\shen}[1]{\todored{shen: #1}}
\newcommand{\response}[1]{\todoblue{#1}}

\begin{abstract}
Code language models have demonstrated strong capabilities across a wide range of code intelligence tasks. While the majority of existing research prioritizes performance improvements on benchmark datasets, few of them have focused on the internal interpretability of models—how specific neurons affect linguistic features such as syntax and semantics, which is critical for model transparency, controllability, and reliability. Although various neuron interpretability techniques have been developed in NLP, directly applying them to source code yields suboptimal results due to the unique characteristics of programming languages, such as their formal structure, hierarchical organization, and executability.
In this work, we empirically investigate the intrinsic mechanisms of code LLMs at the neuron level, aiming to localize both language-specific neurons (\emph{i.e.}, neurons that are selectively responsive to individual programming languages) and concept layers (\emph{i.e.}, feed-forward layers that encode language-agnostic representations of code). 
Our study employs two state-of-the-art models, Llama-3.1-8B and Qwen2.5-Coder-32B, across five programming languages: C++, Java, Python, Go, and JavaScript. By analyzing neuron activation patterns in response to multilingual code inputs, we investigate the role of individual neurons and the contribution of different layers during output generation.
Our empirical findings reveal that: (1) code LLMs contain neurons specialized for individual programming languages, alongside a universal subset that supports general-purpose code generation; and (2) lower layers primarily encode language-specific syntactic structures, while middle layers capture semantic abstractions that generalize across languages, manifesting as concept layers.
To demonstrate the practical usability of these findings, we apply our findings to three downstream tasks: neuron-guided fine-tuning for code generation, clone detection using concept-layer embeddings, and transfer learning guided by concept-layer representations for code summarization. Experimental evaluations show that each strategy consistently improves the performance of multilingual code LLMs.

\end{abstract}



\maketitle

\section{Introduction}

Large language models (LLMs), trained on extensive code repositories, have become powerful productivity tools in software development. They demonstrate strong performance across diverse code-related tasks such as code completion \cite{chaohu2026}, summarization \cite{SunMLZFLDLC25}, translation \cite{chaofan2026}, and code question answering \cite{GLHYHNW0Y24}. 
However, current research on code LLMs primarily focuses on enhancing predictive accuracy across benchmarks, often treating the models as black boxes. For example, approximately 96\% of studies prioritize performance improvements \cite{JiarpakdeeTG21}, while largely neglecting the analysis of internal decision mechanisms. This narrow emphasis leaves the inner workings of code LLMs poorly understood, which may compromise model controllability and reduce trustworthiness~\cite{BlackBox}.

Recently, the interpretability of code LLMs has attracted significant attention from both academic and industrial communities. Nevertheless, substantial challenges remain that hinder a deeper and more reliable understanding of these models' inner workings. 
First, the scalability and computational efficiency of existing explanation methods often fail to keep pace with the rapidly growing size and complexity of modern LLMs. 
Second, there is no established ``ground truth'' for interpretability in code models, making it difficult to objectively evaluate the accuracy and quality of explanatory outputs~\cite{ZhaoCYLDCWYD24}. 
Third, code LLMs exhibit characteristics unique to programming languages (PLs), shaped by code syntax, control flow, and data dependencies \cite{LiuTLL24}. As a result, simply transferring explainability techniques from natural language processing (NLP) frequently leads to suboptimal performance \cite{MohammadkhaniTH23}. For example, the direct use of LAPE~\cite{TangLH0WZWW24} proves ineffective in accurately identifying programming language-specific neurons, as demonstrated in Section~\ref{sec:motivation}.
These challenges demand a novel interdisciplinary approach that integrates deep learning interpretability with principles from programming languages and coding idioms. 
Our goal is to shift the focus from ``which tokens are important'' to ``how internal representations map to and manipulate underlying program structures''.

In this paper, we investigate the internal mechanisms of code LLMs to identify network regions that encode programming language-specific semantic representations and utilize these insights to improve the performance of downstream code-related tasks. 
Specifically, we examine code LLMs at both neuron and layer levels. RQ1 explores the models’ behavior in terms of their internal neuron activations. 
RQ2 examines how the model's output is built through its feed-forward layers and pinpoints potential ``concept layers'' that capture abstract, language-agnostic semantics of code.
Building upon these findings, RQ3 focuses on leveraging these insights to enhance performance on downstream tasks such as code generation, clone detection, and code summarization. 
Our study employs two publicly available LLMs, namely Llama-3.1-8B and Qwen2.5-Coder-32B, along with five popular programming languages: C++, Java, Python, Go, and JavaScript.

We explore the following research questions:

\begin{itemize}[itemsep=1.5mm]

\item \textbf{RQ1: Are there programming language-specific neurons in code LLMs?}

\textbf{Motivation}. Language-specific neurons within a code LLM help to enhance parameter-efficient fine-tuning during adaptation to a specific language by selectively freezing non-specialized neurons.

\textbf{Protocol}. In this research question, we propose the Programming Language Specialization (PLS) score, a novel metric that identifies neurons specialized for individual programming languages. The PLS score evaluates a neuron's activation strength and the gradient of the loss with respect to that activation to identify neurons critical for specific languages. We then validate the language-specificity of these neurons through ablation-based intervention.

\textbf{Result}. Code LLMs contain neurons specialized for individual programming languages as well as a distinct set of universal neurons critical for general code generation. Notably, these neural representations reflect the phylogenetic relationships among programming languages.

\item \textbf{RQ2: Are there language-agnostic concept layers in code LLMs?}

\textbf{Motivation}. The concept layers could act as both a potent semantic hub for code embedding and a bridge for cross-lingual knowledge transfer.

\textbf{Protocol}. In this research question, we employ a dual-pronged analytical framework that triangulates potential concept layers by using both semantic and syntactic probes. We use Representational Similarity Analysis (RSA)~\cite{Kriegeskorte2008} as a semantic probe and an Abstract Syntax Tree (AST) node prediction probe as a syntactic probe, analyzing the layer-wise representational invariance and fidelity of these two properties.

\textbf{Result}. Code LLMs organize knowledge hierarchically, forming concept layers in the middle layers. They exhibit maximum representational invariance to lexical and syntactic changes while simultaneously demonstrating minimum fidelity in encoding fine-grained syntactic structures. 

\item \textbf{RQ3: How can we utilize the interpretability of code LLMs to enhance code-related tasks?}

\textbf{Motivation}. We demonstrate the practical utility of our insights on language-specific neurons (RQ1) and the concept layers (RQ2) across multiple code-related tasks. 

\textbf{Protocol}. In this research question, we leverage the findings from RQ1 and RQ2 to enhance three downstream software engineering tasks: code generation using neuron-guided fine-tuning, clone detection using concept-layer embeddings, and code summarization via concept-layer guided transfer learning.

\textbf{Result}. Our findings provide empirical evidence that leveraging mechanistic insights leads to significant performance improvements across all three tasks. Specifically, neuron-guided fine-tuning outperforms LoRA by mitigating catastrophic forgetting, concept-layer embeddings enable effective zero-shot semantic clone detection, and concept-layer guided transfer learning facilitates superior performance on low-resource code summarization tasks.

\end{itemize}



\section{Background and Motivation}
\label{sec:background_motivation}

\subsection{LLM Interpretability}
\label{sec:XLLM}

LLM Interpretability seeks to uncover the internal mechanisms, decision-making processes, and representational structures of language models, making their behavior more comprehensible, trustworthy, and controllable for humans \cite{ZhaoCYLDCWYD24}. Research in this area has evolved along several interconnected paths.

One central line of research is the probing and analysis of model representations. Probing techniques examine the information encoded within model activations, using methods such as analyzing specific attention heads~\cite{ClarkKLM19}, decoding embedded concepts~\cite{MorrisKSR23}, or applying representation engineering~\cite{abs-2310-01405, abs-2502-19649}. This also includes direct logit lens-style methods that decode activations into vocabulary tokens to understand what information is represented at various layers and positions~\cite{GhandehariounCP24}. These techniques provide insights into the nuanced ways LLMs process and structure information.

Further work delves into a more granular analysis of components, aiming to interpret the function of fundamental units like neurons and attention heads. Studies have successfully mapped individual neurons to human-interpretable concepts~\cite{GurneeNPHTB23, RosaGC23}. For instance, Tang et al.~\cite{TangLH0WZWW24} introduced Language Activation Probability Entropy (LAPE), a metric to identify language-specific neurons by measuring the entropy of their activations across different languages.
Moving beyond individual components, a growing body of research focuses on circuit-level analysis, which examines how distributed groups of neurons and heads interact to perform specific computations. For example, studies have identified circuits responsible for tasks like indirect object identification~\cite{WangVCSS23}. More broadly, this approach can be used for functional localization, such as pinpointing the regions responsible for storing factual knowledge~\cite{DaiDHSCW22}, without necessarily providing a complete explanation of the entire underlying circuit, as demonstrated by Merullo et al.~\cite{MerulloEP24}.

\subsection{Motivation} 
\label{sec:motivation}
While various interpretability techniques have been proposed for general LLMs, directly applying them to code LLMs is often impractical due to the unique characteristics of code in programming languages~\cite{MohammadkhaniTH23}. Unlike natural language, programming languages often share common constructs, such as loops (`for', `while'), which may cause entropy-based measures to misclassify neurons representing these shared concepts as language-agnostic. 
In addition, lexical and sub-token overlaps (\emph{e.g.}, variable names and operators) across languages can bias entropy calculation by existing techniques. 

To verify this challenge, we conduct a preliminary study using LAPE~\cite{TangLH0WZWW24}, an interpretability technique for natural languages, to analyze code LLMs. LAPE identifies a small and variable set of neurons per language, representing only a minor portion of the model’s total capacity. We apply LAPE to Llama-3.1-8B across five programming languages: C++, Java, Python, Go, and JavaScript. As shown in the left part of Table~\ref{tab:LAPE}, the detected language-specific neurons collectively comprise just 0.75\% of all neurons. 
To validate these neurons, we deactivate the subsets associated with each language and assess their impact on code generation. 
Performance is evaluated on the HumanEval-X~\cite{ZhengXZDWXSW0LS23} benchmark with the pass@3 metric. The results, presented in the right part of Table~\ref{tab:LAPE}, show that these ablations do not consistently degrade performance in the intended language and sometimes even enhance performance in others.
This inconsistency suggests that the neurons highlighted by LAPE are not reliably language-specific. 
Supporting this observation, recent work~\cite{Kargaran0YS25} likewise finds that isolating neurons exclusive to a single language is particularly challenging, especially for closely related languages such as Java and C++.

\begin{table*}[t]
\newlength{\myperfwidth}
\settowidth{\myperfwidth}{\scriptsize\texttt{(-100.0\%)}}
\newcommand{\ablationperf}[2]{%
  #1%
  \pgfmathparse{ifthenelse(#2==0, 0, (#1 - #2) / #2 * 100)}%
  \let\change\pgfmathresult
  \def\myperfnum{\pgfmathprintnumber[fixed, zerofill, precision=1, showpos]{\change}\%}%
  \ifdim\change pt > 0pt%
      \def\myperfcontent{\textcolor{teal}{\texttt{(\myperfnum)}}}%
  \else\ifdim\change pt < 0pt%
      \def\myperfcontent{\textcolor{red}{\texttt{(\myperfnum)}}}%
  \else%
      \def\myperfcontent{\texttt{(\myperfnum)}}%
  \fi\fi%
  \makebox[\myperfwidth][r]{\scriptsize\myperfcontent}%
}
\centering
\caption{A Preliminary Study on Llama-3.1-8B by the LAPE Technique. The cell presents the pass@3 scores for code generation after ablating neurons identified by LAPE, and their $\Delta$\ relative to the baseline is shown in parentheses (\textcolor{teal}{↑} / \textcolor{red}{↓}).}
\label{tab:LAPE}
\small
\resizebox{1\textwidth}{!}{
\begin{tabular}{l|r|ccccc}
\toprule
\multirow{2}{*}{\makecell[c]{\textbf{Ablated}\\ \textbf{Neuron Set}}} & \multirow{2}{*}{\textbf{\# Neurons}} & \multicolumn{5}{c}{\textbf{Code Generation Performance on the Target Language}} \\
\cmidrule(lr){3-7}
 & & \textbf{C++ (\%)} & \textbf{Java (\%)} & \textbf{Python (\%)} & \textbf{Go (\%)} & \textbf{JavaScript (\%)} \\
\midrule
\multicolumn{2}{l}{\textbf{Baseline (No Ablation)}} & 47.56 & 49.39 & 46.34 & 34.76 & 40.85 \\
\midrule
\textbf{C++ Specific}
  & 2,549 \scriptsize(0.186\%)
  & \ablationperf{45.85}{47.56} & \ablationperf{46.53}{49.39} & \ablationperf{48.47}{46.34} & \ablationperf{33.86}{34.76} & \ablationperf{43.34}{40.85} \\
\midrule
\textbf{Java Specific}
  & 1,693 \scriptsize(0.124\%)
  & \ablationperf{44.99}{47.56} & \ablationperf{51.51}{49.39} & \ablationperf{47.78}{46.34} & \ablationperf{33.86}{34.76} & \ablationperf{49.51}{40.85} \\
\midrule
\textbf{Python Specific}
  & 2,752 \scriptsize(0.201\%)
  & \ablationperf{39.05}{47.56} & \ablationperf{39.36}{49.39} & \ablationperf{42.77}{46.34} & \ablationperf{29.41}{34.76} & \ablationperf{26.00}{40.85} \\
\midrule
\textbf{Go Specific}
  & 1,260 \scriptsize(0.092\%)
  & \ablationperf{40.76}{47.56} & \ablationperf{45.09}{49.39} & \ablationperf{47.04}{46.34} & \ablationperf{33.86}{34.76} & \ablationperf{48.28}{40.85} \\
\midrule
\textbf{JavaScript Specific}
  & 1,994 \scriptsize(0.146\%)
  & \ablationperf{44.18}{47.56} & \ablationperf{44.40}{49.39} & \ablationperf{41.38}{46.34} & \ablationperf{31.18}{34.76} & \ablationperf{47.06}{40.85} \\
\midrule
\textbf{Language Agnostic}
  & 14,304 \scriptsize(1.044\%)
  & \textbf{\ablationperf{0.00}{47.56}} & \textbf{\ablationperf{0.00}{49.39}} & \textbf{\ablationperf{0.00}{46.34}} & \textbf{\ablationperf{0.00}{34.76}} & \textbf{\ablationperf{0.00}{40.85}} \\
\bottomrule
\end{tabular}
}
\end{table*}

These factors motivate our study methodology to explain code LLMs by integrating deep learning interpretability with the inherent properties of programming languages.

\section{Experimental Setup} 
\label{sec:setup}

We investigate the intrinsic mechanisms of code LLMs to understand how individual neurons influence linguistic features.
In this section, we discuss the models and programming languages used for our experiments.

\subsection{Language Models under Study} \label{subsec:model}
Our investigation is conducted on two publicly available LLMs: \textbf{Llama-3.1-8B}~\cite{Llama3} and \textbf{Qwen2.5-Coder-32B}~\cite{abs-2409-12186}. Llama-3.1 is a foundational model pre-trained on both broad text and code corpora, whereas Qwen2.5-Coder is a code-specialized model trained exclusively on over 3 trillion high-quality code tokens. Both are autoregressive and decoder-only transformers.
We use the 8B version of Llama-3.1 ($\sim$1.5M neurons) as a lightweight backbone suitable for fine-tuning. For Qwen2.5-Coder, we select the 32B variant ($\sim$4.8M neurons) to examine a larger-scale model. 
We utilize only the base versions of each model, avoiding confounding effects from post-training alignment techniques such as RLHF. This enables a direct and unbiased analysis of the knowledge and architectural properties intrinsically acquired during pre-training.

\subsection{Selected Programming Languages} \label{subsec:languages}
Our study examines five widely adopted programming languages, including C++, Java, Python, Go, and JavaScript, selected for their significant industry presence and distinct linguistic features. The set includes both statically-typed (C++, Java, Go) and dynamically-typed (Python, JavaScript) languages, supporting various paradigms such as object-oriented, procedural, and scripting. It also spans different execution models, from compiled languages with manual memory management (C++) to interpreted or JIT-compiled languages with garbage collection (Python, JavaScript, Go). This diversity provides a comprehensive foundation for evaluating how code LLMs comprehend, distinguish, and generalize across different programming languages.

\subsection{Implementation Details}
\label{subsec:implementation}
All experiments are conducted within a standardized software and hardware environment to ensure reproducibility. We utilize the HuggingFace transformers~\cite{WolfDSCDMCRLFDS20} (v4.43.0) and PyTorch~\cite{Paszke2019PyTorch} (v2.3.0) libraries. The experimental platform is a workstation equipped with two NVIDIA GeForce RTX 4090 GPUs, running on CUDA 12.2. All base models are loaded directly from the Hugging Face Hub. To optimize for computational efficiency and memory usage, all training and inference operations are performed using the \texttt{torch.bfloat16} mixed-precision format.

For all fine-tuning tasks, we adopt a consistent training configuration unless otherwise specified. We employ the AdamW optimizer~\cite{LoshchilovH19} with a learning rate of $5 \times 10^{-5}$ and a linear scheduler featuring a 10\% warm-up phase. We use a maximum sequence length of 1,024 tokens and a per-device batch size of 4, with gradient accumulation over 2 steps to achieve an effective batch size of 16. A fixed random seed of 42 is used across all experiments to ensure deterministic outcomes. The specific hyperparameters for each downstream task, such as the number of epochs and the exact fine-tuning strategy, are detailed in Section \ref{sec:rq3}.

\section{RQ1: Identifying Programming Language-Specific Neurons}
\label{sec:rq1}

In this study, we investigate neuron activations in response to multilingual code inputs and analyze the distribution of programming language-specific neurons in code LLMs.

\subsection{Study Design}
\label{sec:methodology_rq1}

\begin{figure}[t!]
    \centering 
    \includegraphics[width=0.95\linewidth]{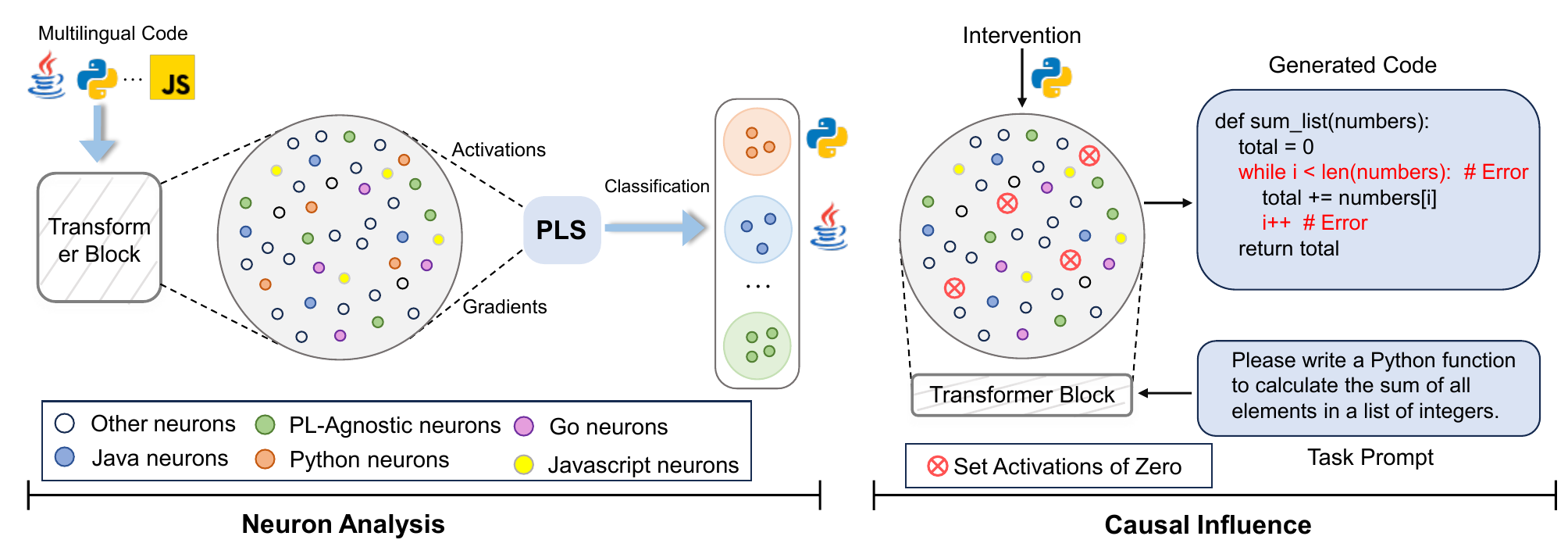} 
    \caption{The Overall Structure of Our Analysis Framework for Language-Specific Neurons.}
    \label{fig:rq1_methodology}
\end{figure}

To address the limitations of NL-oriented measures like LAPE, we introduce the Programming Language Specialization (PLS) score, a novel metric that identifies programming language-specific neurons. The PLS evaluates both a neuron’s activation strength and the gradient of the loss with respect to its activation, distinguishing neurons activated by linguistic features from those that are functionally essential for accurate language generation. 
The analysis framework, shown in Figure~\ref{fig:rq1_methodology}, consists of two main stages: (1) identifying language-specific neurons using PLS scoring, and (2) evaluating the influence of these neurons by ablating their activations and measuring the resulting change in model performance. Each stage is described in detail in the following sections.

\subsubsection{Probing Language-Specific Neurons}
\label{sec:identifying}
We employ gradient-based attribution methods~\cite{SimonyanVZ13, SundararajanTY17} to evaluate a feature's importance by multiplying its activation with the loss gradient. Extending this approach to internal neuron activations, we quantify each neuron’s contribution to model predictions.

For a neuron \( n \) and an input code sample \( s \), let \( a_n(s) \) denote its activation. The gradient of the sequence-level cross-entropy loss \( \mathcal{L}(s) \) with respect to this activation is \( \nabla_{a_n(s)} \mathcal{L}(s) \), where the loss is defined as:
\begin{equation}
\mathcal{L}(s) = - \sum_{t=1}^{T} \log p_\theta(x_t \mid x_{<t}),
\end{equation}
with \( (x_1, \dots, x_T) \) representing the ground-truth token sequence and \( p_\theta \) the model’s predictive distribution.

The PLS score of neuron \( n \) for language \( L \) is defined as the expectation over language-specific samples \( s \in D_L \) of the product of the absolute activation and the absolute gradient:
\begin{equation}
\text{PLS}(n, L) = \mathbb{E}_{s \in D_L} \left[ |a_n(s)| \cdot \left| \nabla_{a_n(s)} \mathcal{L}(s) \right| \right],
\end{equation}
where the expectation is taken over code samples \( s \) in the language-specific dataset \( D_L \).

Language-specific neurons are identified through the following three-step procedure:

\begin{enumerate}
    \item \textit{Candidate Filtering.} A neuron \( n \) is considered a candidate if its PLS score exceeds a predefined threshold in at least one language. Neurons with uniformly low scores are discarded.
    
    \item \textit{Specificity Verification.} For each candidate, let \( L^\ast \) be the language with the highest PLS score:
    \begin{equation}
    L^\ast = \arg\max_{L_i \in \mathbb{L}} \text{PLS}(n, L_i).
    \end{equation}
    A neuron is considered language-specific if this score is at least 1.5 times greater than the second-highest score, balancing precision and coverage.
    
    \item \textit{Ranking and Selection.} Neurons are ranked by their maximal PLS score, and the top 1\% are classified as \emph{language-specialized}, each attributed to its corresponding \( L^\ast \).
\end{enumerate}
This procedure yields a final set of neurons that are both highly responsive and influential for the model’s predictions, while being clearly specific to individual programming languages.

\subsubsection{Validating Language-Specific Neurons}
\label{sec:validation_protocol}
We adopt a standard ablation-based intervention protocol~\cite{DaiDHSCW22} to validate the identified language-specific neurons in code generation. It involves perturbing neuron activations and assessing the impact on the quality of the generated code:

\begin{enumerate}
    \item \textit{Intervention:} During the forward pass for a code generation task, we ablate a previously identified set of specialized neurons (\emph{e.g.}, those for Python) by setting their activation values to zero immediately after computation. This intervention, applied within their respective neural modules (both self-attention and feed-forward layers), nullifies their contribution to subsequent layers and the final output.

    \item \textit{Evaluation:} We evaluate the model's performance on the HumanEval-X~\cite{ZhengXZDWXSW0LS23} benchmark, a multilingual dataset for assessing cross-lingual code synthesis. For each problem, the model generates three candidate solutions, and performance is quantified using the \textbf{pass@3} metric, which measures the probability that at least one of the three solutions passes all unit tests.
\end{enumerate}

We compare the results of ablation guided by our proposed PLS scores with random ablation using an equal number of neurons. A sharp performance decline in the target language, with minimal impact on other languages, suggests high accuracy in identifying language-specific neurons.

\subsection{Results}
\label{sec:rq1_results}

We apply the PLS technique to the studied models and identify specialized sets of neurons for C++, Java, Python, Go, and JavaScript, along with a language-agnostic neuron set that exhibits high activation across all languages. The distribution of language-specific neurons is summarized in Table~\ref{tab:rq1_intervention_results}. Notably, these language-specific neurons account for less than 0.7\% of the total in Llama-3.1-8B and 0.5\% in Qwen2.5-Coder-32B. The results also reveal a significant quantitative imbalance across languages. For example, in Llama-3.1-8B, the number of Python-specific neurons (4,656) is more than 14 times greater than that of Go-specific neurons (315). This imbalance occurs because PLS relies on aggregated activation values and their gradient-based attribution. Languages that trigger stronger model activations, such as Python, receive more highly attributed neurons.

\begin{table*}[t]
\settowidth{\myperfwidth}{\scriptsize\texttt{(-100.0\%)}}
\newcommand{\ablationperf}[2]{%
  #1%
  \pgfmathparse{ifthenelse(#2==0, 0, (#1 - #2) / #2 * 100)}%
  \let\change\pgfmathresult
  \def\myperfnum{\pgfmathprintnumber[fixed, zerofill, precision=1, showpos]{\change}\%}%
  \ifdim\change pt > 0pt%
      \def\myperfcontent{\textcolor{teal}{\texttt{(\myperfnum)}}}%
  \else\ifdim\change pt < 0pt%
      \def\myperfcontent{\textcolor{red}{\texttt{(\myperfnum)}}}%
  \else%
      \def\myperfcontent{\texttt{(\myperfnum)}}%
  \fi\fi%
  \makebox[\myperfwidth][r]{\scriptsize\myperfcontent}%
}

\centering
\caption{Code Generation Performance after Ablating PLS-Identified Neurons (HumanEval-X). The percentage change ($\Delta$\%) relative to the baseline is shown in parentheses (\textcolor{teal}{↑} / \textcolor{red}{↓}).}
\label{tab:rq1_intervention_results}
\small
\resizebox{1\textwidth}{!}{
\begin{tabular}{l|r|l|ccccc}
\toprule
\multirow{2}{*}{\makecell[c]{\textbf{Ablated}\\ \textbf{Neuron Set}}} & \multirow{2}{*}{\textbf{\# Neurons}} & \multirow{2}{*}{\makecell[c]{\textbf{Ablating}\\\textbf{Method}}} & \multicolumn{5}{c}{\textbf{Code Generation Performance on the Target Language}} \\
\cmidrule(lr){4-8}
 & & & \textbf{C++ (\%)} & \textbf{Java (\%)} & \textbf{Python (\%)} & \textbf{Go (\%)} & \textbf{JavaScript (\%)} \\
\midrule
\multicolumn{8}{>{\columncolor{gray!20}}c}{\textbf{\textsc{Llama-3.1-8B}}} \\
\midrule
\multicolumn{2}{l|}{\textbf{Baseline (No Ablation)}} & - & 47.56 & 49.39 & 46.34 & 34.76 & 40.85 \\
\midrule
\multirow{2}{*}{\textbf{C++ Specific}}
 & \multirow{2}{*}{2,290 \scriptsize(0.167\%)} & PLS    & \textbf{\ablationperf{6.71}{47.56}} & \ablationperf{31.71}{49.39} & \ablationperf{48.17}{46.34} & \ablationperf{25.00}{34.76} & \ablationperf{31.10}{40.85} \\
 & & Random & \ablationperf{31.10}{47.56} & \ablationperf{42.07}{49.39} & \ablationperf{36.59}{46.34} & \ablationperf{24.39}{34.76} & \ablationperf{25.00}{40.85} \\
\midrule
\multirow{2}{*}{\textbf{Java Specific}}
 & \multirow{2}{*}{1,008 \scriptsize(0.074\%)} & PLS    & \ablationperf{43.29}{47.56} & \textbf{\ablationperf{27.44}{49.39}} & \ablationperf{45.12}{46.34} & \ablationperf{35.32}{34.76} & \ablationperf{45.73}{40.85} \\
 & & Random & \ablationperf{35.98}{47.56} & \ablationperf{51.22}{49.39} & \ablationperf{45.73}{46.34} & \ablationperf{27.44}{34.76} & \ablationperf{36.59}{40.85} \\
\midrule
\multirow{2}{*}{\textbf{Python Specific}}
 & \multirow{2}{*}{4,656 \scriptsize(0.340\%)} & PLS    & \ablationperf{29.26}{47.56} & \ablationperf{36.58}{49.39} & \textbf{\ablationperf{7.93}{46.34}} & \ablationperf{26.83}{34.76} & \ablationperf{9.15}{40.85} \\
 & & Random & \ablationperf{36.59}{47.56} & \ablationperf{40.24}{49.39} & \ablationperf{37.80}{46.34} & \ablationperf{31.10}{34.76} & \ablationperf{37.20}{40.85} \\
\midrule
\multirow{2}{*}{\textbf{Go Specific}}
 & \multirow{2}{*}{315 \scriptsize(0.023\%)} & PLS    & \ablationperf{38.41}{47.56} & \ablationperf{49.39}{49.39} & \ablationperf{44.51}{46.34} & \textbf{\ablationperf{4.27}{34.76}} & \ablationperf{43.90}{40.85} \\
 & & Random & \ablationperf{42.68}{47.56} & \ablationperf{53.66}{49.39} & \ablationperf{48.17}{46.34} & \ablationperf{32.93}{34.76} & \ablationperf{42.68}{40.85} \\
\midrule
\multirow{2}{*}{\textbf{JavaScript Specific}}
 & \multirow{2}{*}{1,148 \scriptsize(0.084\%)} & PLS    & \ablationperf{41.46}{47.56} & \ablationperf{50.61}{49.39} & \ablationperf{50.00}{46.34} & \ablationperf{35.98}{34.76} & \textbf{\ablationperf{23.17}{40.85}} \\
 & & Random & \ablationperf{45.12}{47.56} & \ablationperf{52.44}{49.39} & \ablationperf{46.95}{46.34} & \ablationperf{36.59}{34.76} & \ablationperf{40.24}{40.85} \\
\midrule
\multirow{2}{*}{\textbf{Language Agnostic}}
 & \multirow{2}{*}{12,087 \scriptsize(0.882\%)} & PLS    & \textbf{\ablationperf{0.00}{47.56}} & \textbf{\ablationperf{0.00}{49.39}} & \textbf{\ablationperf{0.00}{46.34}} & \textbf{\ablationperf{0.00}{34.76}} & \textbf{\ablationperf{0.00}{40.85}} \\
 & & Random & \ablationperf{14.64}{47.56} & \ablationperf{16.47}{49.39} & \ablationperf{21.34}{46.34} & \ablationperf{7.95}{34.76} & \ablationperf{12.36}{40.85} \\
\midrule
\multicolumn{8}{>{\columncolor{gray!20}}c}{\textbf{\textsc{Qwen2.5-Coder-32B}}} \\
\midrule
\multicolumn{2}{l|}{\textbf{Baseline (No Ablation)}} & - & 79.27 & 76.83 & 76.83 & 71.34 & 98.17 \\
\midrule
\multirow{2}{*}{\textbf{C++ Specific}}
 & \multirow{2}{*}{1,931 \scriptsize(0.042\%)} & PLS    & \textbf{\ablationperf{40.24}{79.27}} & \ablationperf{58.54}{76.83} & \ablationperf{68.90}{76.83} & \ablationperf{59.15}{71.34} & \ablationperf{85.37}{98.17} \\
 & & Random & \ablationperf{73.17}{79.27} & \ablationperf{76.22}{76.83} & \ablationperf{77.44}{76.83} & \ablationperf{70.73}{71.34} & \ablationperf{97.56}{98.17} \\
\midrule
\multirow{2}{*}{\textbf{Java Specific}}
 & \multirow{2}{*}{1,809 \scriptsize(0.039\%)} & PLS    & \ablationperf{74.39}{79.27} & \textbf{\ablationperf{40.24}{76.83}} & \ablationperf{73.17}{76.83} & \ablationperf{62.20}{71.34} & \ablationperf{98.17}{98.17} \\
 & & Random & \ablationperf{79.88}{79.27} & \ablationperf{71.95}{76.83} & \ablationperf{75.61}{76.83} & \ablationperf{71.95}{71.34} & \ablationperf{97.56}{98.17} \\
\midrule
\multirow{2}{*}{\textbf{Python Specific}}
 & \multirow{2}{*}{9,400 \scriptsize(0.205\%)} & PLS    & \ablationperf{75.00}{79.27} & \ablationperf{60.78}{76.83} & \textbf{\ablationperf{43.90}{76.83}} & \ablationperf{68.29}{71.34} & \ablationperf{96.95}{98.17}\\
 & & Random & \ablationperf{64.02}{79.27} & \ablationperf{59.76}{76.83} & \ablationperf{54.88}{76.83} & \ablationperf{51.83}{71.34} & \ablationperf{82.32}{98.17} \\
\midrule
\multirow{2}{*}{\textbf{Go Specific}}
 & \multirow{2}{*}{6,110 \scriptsize(0.133\%)} & PLS    & \ablationperf{75.61}{79.27} & \ablationperf{66.46}{76.83} & \ablationperf{71.34}{76.83} & \textbf{\ablationperf{4.27}{71.34}} & \ablationperf{92.07}{98.17} \\
 & & Random & \ablationperf{70.12}{79.27} & \ablationperf{65.85}{76.83} & \ablationperf{67.07}{76.83} & \ablationperf{57.98}{71.34} & \ablationperf{88.41}{98.17} \\
\midrule
\multirow{2}{*}{\textbf{JavaScript Specific}}
 & \multirow{2}{*}{1,463 \scriptsize(0.032\%)} & PLS    & \ablationperf{71.34}{79.27} & \ablationperf{74.39}{76.83} & \ablationperf{67.68}{76.83} & \ablationperf{70.12}{71.34} & \textbf{\ablationperf{79.87}{98.17}} \\
 & & Random & \ablationperf{78.66}{79.27} & \ablationperf{77.44}{76.83} & \ablationperf{76.83}{76.83} & \ablationperf{70.73}{71.34} & \ablationperf{94.51}{98.17} \\
\midrule
\multirow{2}{*}{\textbf{Language Agnostic}}
 & \multirow{2}{*}{27,086 \scriptsize(0.589\%)} & PLS    & \textbf{\ablationperf{0.00}{79.27}} & \textbf{\ablationperf{0.00}{76.83}} & \textbf{\ablationperf{0.00}{76.83}} & \textbf{\ablationperf{0.00}{71.34}} & \textbf{\ablationperf{0.00}{98.17}} \\
 & & Random & \ablationperf{25.61}{79.27} & \ablationperf{28.05}{76.83} & \ablationperf{30.49}{76.83} & \ablationperf{21.95}{71.34} & \ablationperf{50.00}{98.17} \\ 
\bottomrule
\end{tabular}
}
\end{table*}

Table~\ref{tab:rq1_intervention_results} presents the code generation performance after deactivating the language-specific neurons identified by PLS. A sharp performance decline is observed in the ablated language, with minimal impact on others. Compared to LAPE (Table~\ref{tab:LAPE}), PLS demonstrates higher accuracy in pinpointing language-specific neurons. For instance, in Llama-3.1-8B, ablating Python-specific neurons identified by PLS leads to a performance drop of 82.9\%, whereas LAPE-based ablation results in only a 7.7\% decrease. The gap is even more pronounced for C++: PLS-driven ablation causes an 85.9\% decline, compared to a negligible 3.6\% decrease with LAPE. These results confirm the importance of causal-based analysis, demonstrating that purely correlation-driven metrics like LAPE are insufficient for this task.

Additionally, the PLS method demonstrates both computational efficiency and scalability, yielding consistent results across models of varying sizes, such as the smaller Llama-3.1-8B and the larger Qwen2.5-Coder-32B. Despite the substantial difference in model size, PLS remains effective in identifying language-specific neurons, indicating its ability to handle the complexities inherent in models of different scales.

Beyond superior accuracy, the neurons identified by PLS also exhibit remarkable functional specificity. For instance, ablating the Go-specific neuron set induces a performance collapse of 87.8\% (Llama-3.1-8B) and 94.0\% (Qwen2.5-Coder-32B) for Go (\emph{e.g.}, from 34.76\% to 4.27\% in Llama-3.1-8B), yet has a negligible impact on Python's performance. This targeted effect stands in stark contrast to the generalized degradation from random ablation. Furthermore, ablating the language-agnostic neuron set incapacitates the model entirely, reducing performance to 0.00\% across all languages and confirming their foundational role in encoding universal programming constructs.

Ablation experiments reveal complex patterns of collateral performance degradation, often corresponding to the phylogenetic relationships between languages. A consistent, though asymmetric, relationship is observed between C++ and Java, which share C-style syntax. Ablating C++-specific neurons results in a significant performance drop in Java, with reductions of 35.8\% for Llama-3.1-8B and 23.6\% for Qwen2.5-Coder-32B. This indicates that both models capture a shared neural representation reflecting the syntactic similarities between these languages, with C++-specific neurons having a more pronounced effect on Java performance.

A more pronounced, yet model-dependent, interaction is observed between Python and JavaScript, two dynamically-typed scripting languages. For Llama-3.1-8B, ablating Python-specific neurons results in a substantial 77.6\% decline in JavaScript performance, whereas for Qwen2.5-Coder-32B, the impact remains negligible. This stark divergence suggests that the models encode these languages in fundamentally different ways. Moreover, in \mbox{Llama-3.1-8B}, this relationship is further complicated as ablating JavaScript-specific neurons unexpectedly enhances Python performance, implying a complex inhibitory or competitive dynamic rather than a simple feature-sharing mechanism. These discrepancies are likely attributable to variations in model architectures or the different emphasis placed on certain aspects of the training data.

\begin{tcolorbox}[width=\linewidth, boxrule=0.8pt, left=2pt, right=2pt, top=2pt, bottom=2pt, colback=gray!10, colframe=black]
\textbf{Finding 1:} Code LLMs possess a small proportion of neurons specialized for individual programming languages (less than 0.7\%), as well as a language-agnostic set (less than 0.9\%) that exhibits strong causal influence across all languages.
\end{tcolorbox}
\section{RQ2: Characterizing Layer-wise Functional Specialization} \label{sec:rq2}


Having identified language-specific neurons, we further investigate whether code LLMs exhibit language-agnostic ``concept layers'' that encode abstract, language-agnostic representations of code. These identified concept layers could serve as potent semantic hubs for code embedding and a bridge for cross-lingual knowledge transfer. We focus on examining the feed-forward layers of the Transformer model. Previous studies in NLP~\cite{AkenWLG19, JawaharSS19, TenneyDP19} have shown that feed-forward networks (FFNs) function as key memory components in LLMs, developing a feature hierarchy that transitions from surface-level syntax in lower layers to more abstract semantics in higher layers.

\subsection{Study Design}

We introduce a dual-pronged analytical framework to identify candidate concept layers using complementary semantic and syntactic probes. First, we apply Representational Similarity Analysis (RSA)~\cite{Kriegeskorte2008} as a semantic probe to measure the invariance of layer-wise representations to both lexical variations and syntactic changes, a key indicator of abstraction. Second, we deploy a syntactic probe based on AST node prediction to track how grammatical structures are encoded across layers. The expectation is that a genuine concept layer should exhibit high semantic invariance while simultaneously demonstrating reduced sensitivity to fine-grained syntactic details.

\subsubsection{Probing Semantic Abstraction}

We apply RSA to investigate semantic abstraction in code LLMs on the HumanEval-X~\cite{ZhengXZDWXSW0LS23} benchmark. RSA enables the identification of \textit{concept layers} by analyzing the representational geometry across model layers. The core premise is that a concept layer should capture semantic equivalence: code snippets with the same functionality should map to similar representations, even if they differ in surface form (e.g., variable names or syntactic structures).

To probe this property, we design two controlled experiments targeting distinct types of code modification. By examining whether certain layers exhibit consistent representations under these variations, we can identify which layers serve as candidates for semantic abstraction in code LLMs.

The first experiment tests intra-lingual invariance to lexical changes, evaluating whether the model remains unaffected by superficial identifier alterations within a single programming language. The ability to ignore such trivial differences is essential for genuine semantic understanding. If a representation changes significantly due solely to variable renaming, it suggests a failure to capture the underlying algorithmic logic. For each original function \( F_{\text{orig}} \), we generate a semantically equivalent version \( F_{\text{anon}} \) by replacing all user-defined identifiers with random strings. Concretely, we only rename user-defined variables (e.g., local variables, parameters, and fields), leaving language keywords and library symbols unchanged. Each identifier is substituted with a short random alphanumeric string of length 2–7 to avoid introducing out-of-distribution naming patterns. After renaming, we execute the original HumanEval-X unit tests and retain only functions that pass all test cases, ensuring that \( F_{\text{orig}} \) and \( F_{\text{anon}} \) are both syntactically valid and behaviorally equivalent. We then measure the representational similarity between each \( (F_{\text{orig}}, F_{\text{anon}}) \) pair.

The second experiment evaluates cross-lingual equivalence under syntactic changes, directly probing the existence of language-agnostic concept layers by assessing representational invariance across major syntactic variations. By measuring the similarity between semantically equivalent functions written in different languages (\emph{e.g.}, Python and Java), we gauge the model’s capacity to look beyond divergent grammars, keywords, and type systems. High similarity here would strongly indicate that the model captures an abstract algorithmic core, independent of its syntactic expression. We use the parallel corpus of the HumanEval-X benchmark to compare function pairs that solve the same problem in different languages (\emph{e.g.}, \( (F_{\text{python}}, F_{\text{java}}) \)).

For both experiments, the following steps are performed to quantify layer-wise representational similarity:

\begin{enumerate}
    \item \textit{Representation Extraction:} Each code snippet is processed by a frozen model. For each transformer layer, we extract the final hidden-state vector for every input token.
    
    \item \textit{Embedding Aggregation:} To obtain a single vector representation of the entire code snippet at a specific layer, mean pooling is applied over the sequence of token vectors. This standard aggregation method produces a fixed-size embedding $V_l(s)$ for a snippet $s$ at layer $l$.
    
    \item \textit{Similarity Quantification:} We compute the similarity between the representations of two code snippets, $s_A$ and $s_B$, at each layer using cosine similarity, which measures the angular closeness between embedding vectors. For a given pair, the layer-wise similarity is defined as:
    \begin{equation} \label{eq:sim_consine}
    Sim(V_l(s_A), V_l(s_B)) = \frac{V_l(s_A) \cdot V_l(s_B)}{|V_l(s_A)| |V_l(s_B)|}.
    \end{equation}
    These similarity scores are averaged across all code pairs. The resulting intra-lingual invariance score quantifies the model's robustness to superficial lexical variations, while the cross-lingual equivalence score evaluates the degree of abstraction from specific syntactic structures. Together, these metrics provide a quantitative foundation for assessing the formation of semantic concept layers across the model’s depth.
\end{enumerate}

\subsubsection{Probing Syntactic Structure}
Along with the semantic probe, a syntactic probe is deployed to map the layer-wise encoding of grammatical structure. This analysis delineates the boundaries of potential concept layers by identifying where the model's focus shifts from syntactic processing to semantic abstraction. To this end, we adapt established NLP probing techniques, using supervised classifiers to detect syntactic structures by predicting linguistic properties from hidden states~\cite{HewittM19}. Specifically, we use a \textit{linear} classifier to ensure that the probe primarily surfaces information encoded in the representations, rather than relying on its capacity to learn the task. Prior work~\cite{HewittL19, PimentelSWC20} shows that high classification accuracy with a linear probe indicates that syntactic information is linearly separable in the LLM's representation space, demonstrating the model’s ability to organize such structure.

The following steps are performed to quantify the degree to which syntactic information is explicitly encoded at each layer:

\begin{enumerate}
    \item \textit{Data Preparation:} Each code sample from the HumanEval-X dataset is parsed into its corresponding AST.
    
    \item \textit{Token-to-Node Alignment:} A labeled dataset is constructed by identifying the corresponding AST node for each token in an input snippet. This step creates training instances where the input feature is the token's hidden-state vector from a given layer, and the target label is its associated AST node type (\emph{e.g.}, \texttt{FunctionDef}, \texttt{ForStmt}, \texttt{BinOp}).
    
    \item \textit{Layer-wise Probe Training:} For each transformer layer of the frozen model, a linear classifier (multinomial logistic regression) is trained. The underlying model weights are held constant; only the probe's parameters are updated. The probe aims to predict the AST node type from a single token's hidden-state vector at that layer.
    
    \item \textit{Evaluation:} The performance of each layer-specific probe is evaluated using classification accuracy on a held-out test set. This step yields an accuracy score for each layer, generating a curve that illustrates the prominence of syntactic information across the model's depth.
\end{enumerate}

\subsection{Results}

\begin{figure}[t!]
    \centering
    \begin{subfigure}[b]{0.48\linewidth}
        \centering
        \includegraphics[width=\linewidth]{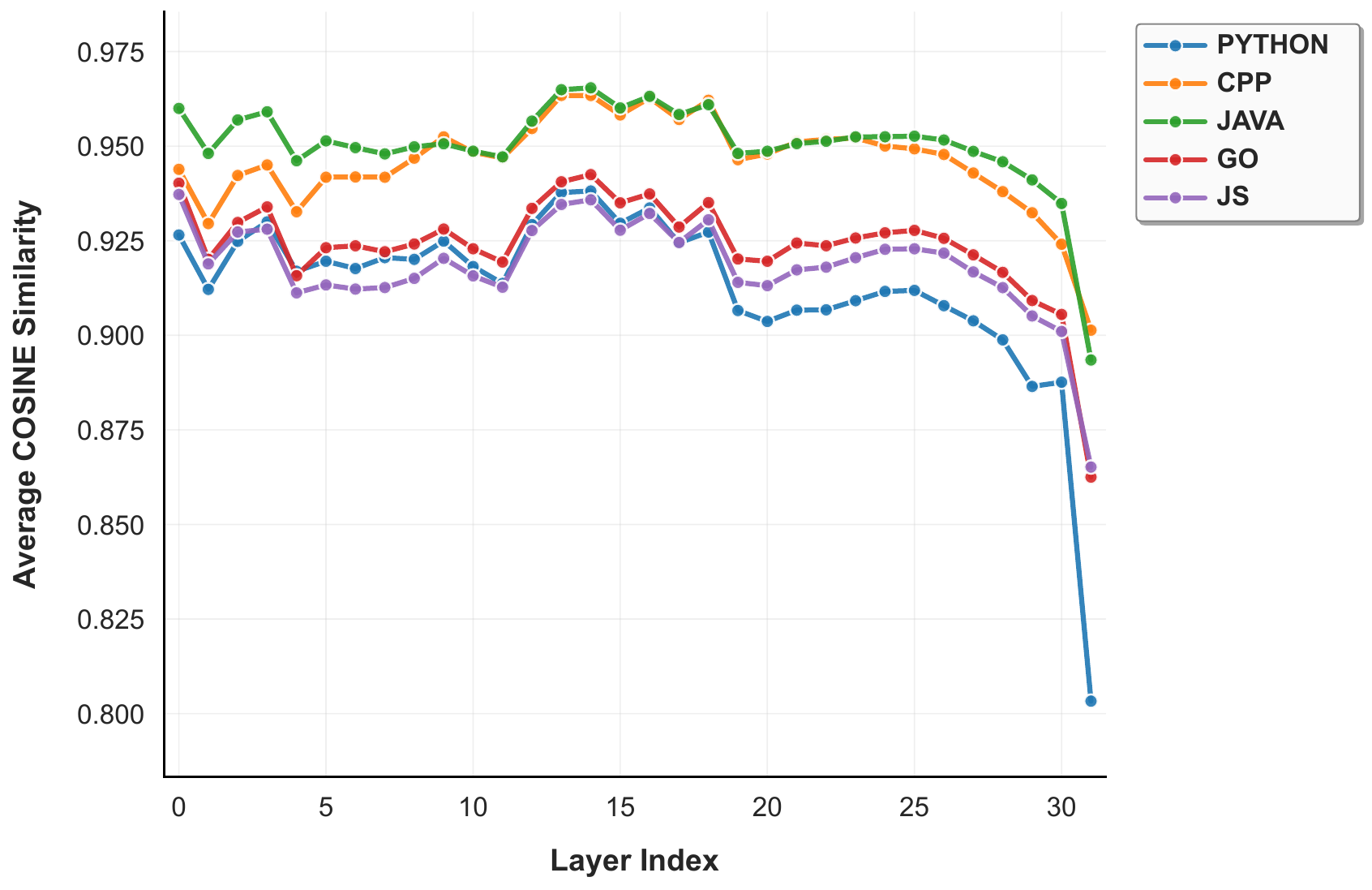}
        \caption{Llama-3.1-8B}
        \label{fig:intra_llama_separate}
    \end{subfigure}
    \hfill 
    \begin{subfigure}[b]{0.48\linewidth}
        \centering
        \includegraphics[width=\linewidth]{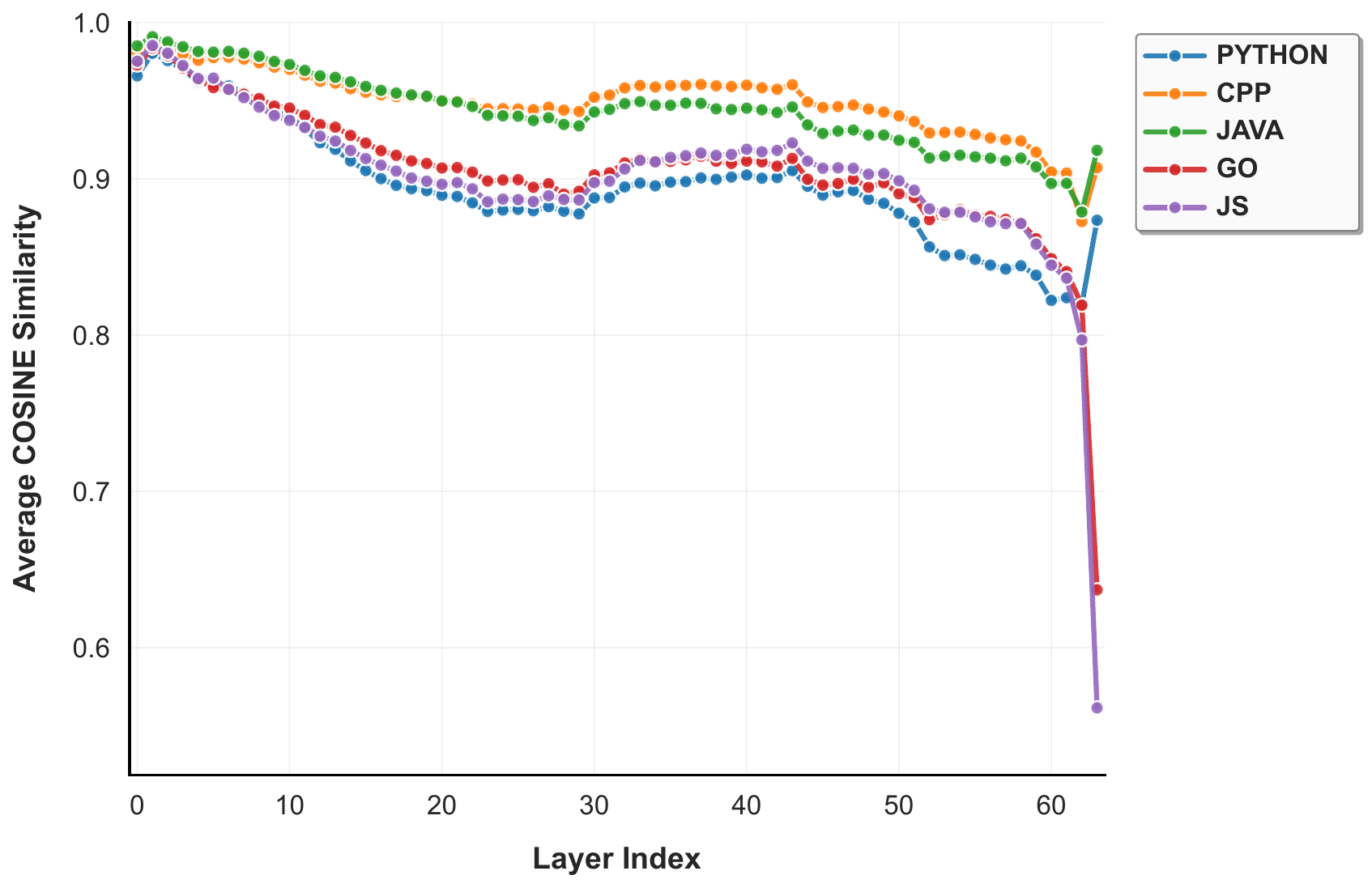}
        \caption{Qwen2.5-Coder-32B}
        \label{fig:intra_qwen_separate}
    \end{subfigure}
    \caption{
        Intra-lingual Semantic Invariance Under Lexical Variations across Model Layers. 
    }
    \label{fig:intra_lingual_similarity}
\end{figure}

\smallskip

\begin{figure}[t!]
    \centering
    \begin{subfigure}[b]{0.48\linewidth}
        \centering
        \includegraphics[width=\linewidth]{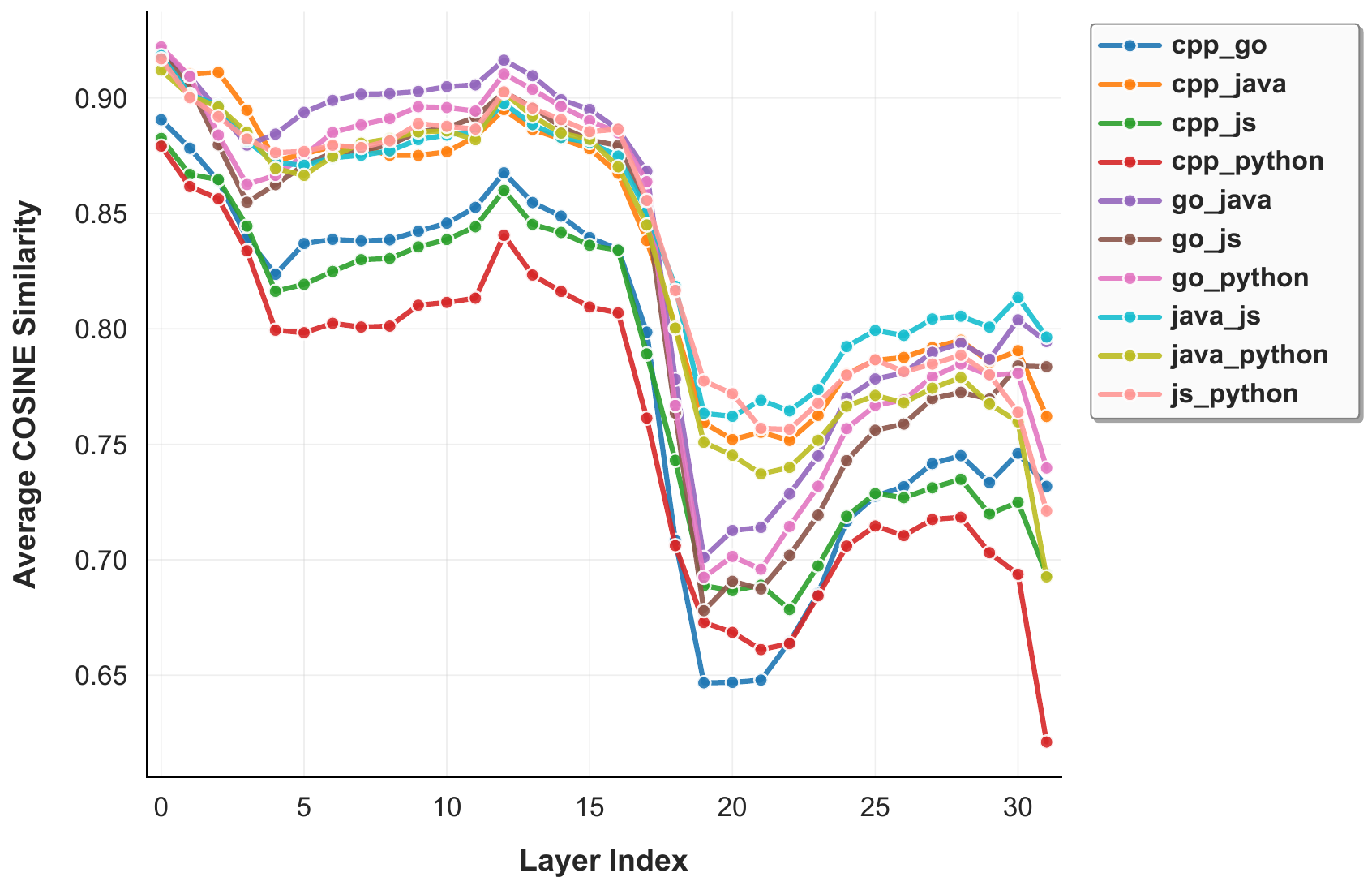}
        \caption{Llama-3.1-8B}
        \label{fig:cross_llama_separate}
    \end{subfigure}
    \hfill
    \begin{subfigure}[b]{0.48\linewidth}
        \centering
        \includegraphics[width=\linewidth]{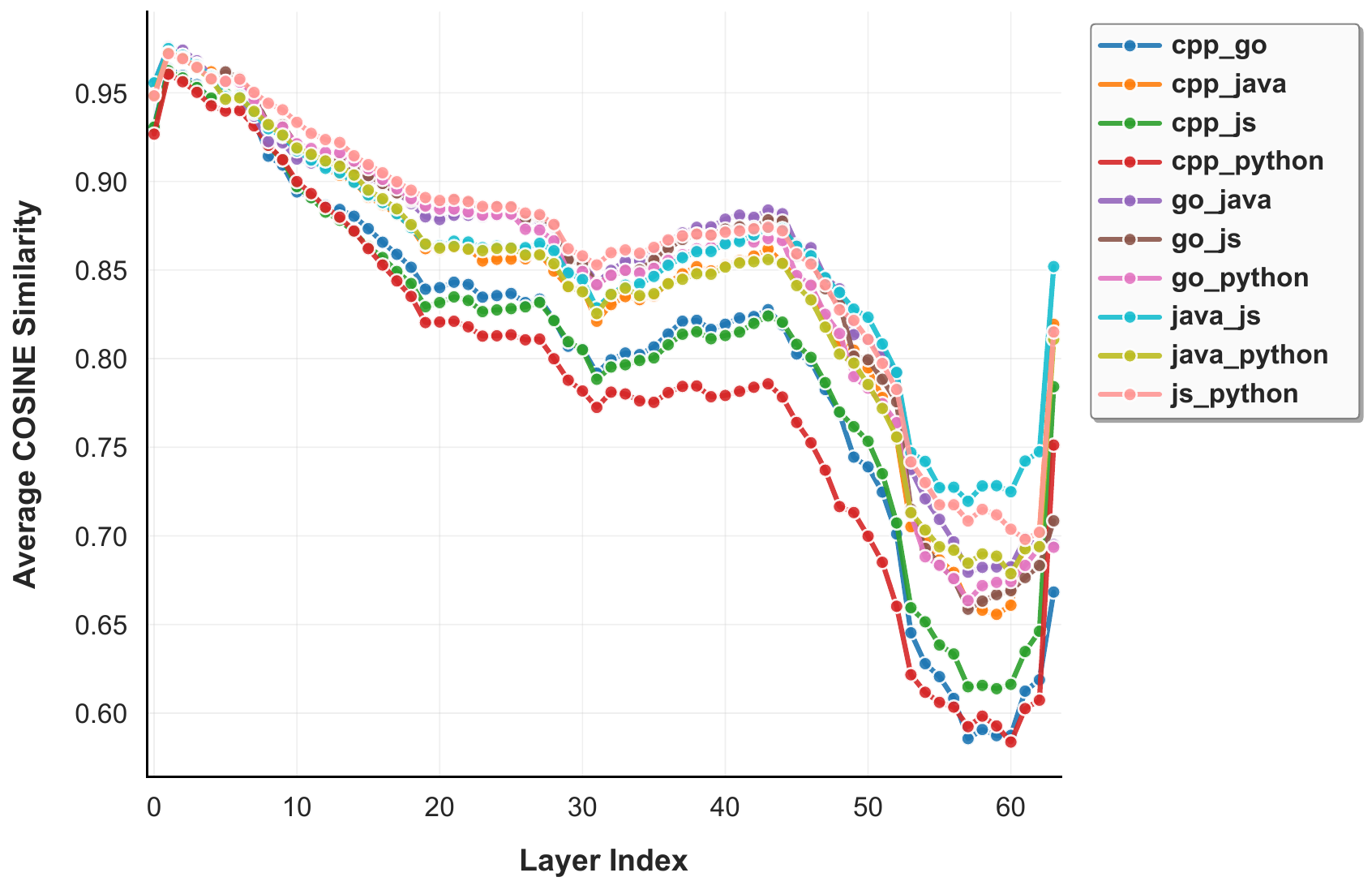}
        \caption{Qwen2.5-Coder-32B}
        \label{fig:cross_qwen_separate}
    \end{subfigure}
    \caption{
        Cross-lingual Semantic Equivalence under Syntactic Variations across Model Layers. 
    }
    \label{fig:cross_lingual_similarity}
\end{figure}


We apply both a semantic probe with RSA and a syntactic probe with AST node prediction to measure how invariant layer-wise representations are to lexical and syntactic changes. We refer to \emph{concept layers} as regions of layers that jointly satisfy two criteria: (i) they exhibit high semantic invariance under both intra-lingual and cross-lingual transformations, and (ii) they have low syntactic decodability, meaning that fine-grained syntactic details are no longer linearly recoverable. We next summarize how these two probes jointly reveal such layers.

\textbf{Results of Semantic Abstraction Probing}. Figure~\ref{fig:intra_lingual_similarity} shows intra-lingual invariance under lexical variations, measured by the cosine similarity between original and variable-renamed code representations. In Llama-3.1-8B, similarity is highest in a band of middle layers roughly spanning layers 8–15, while in Qwen2.5-Coder-32B it remains consistently high across layers 31–43. These bands indicate that the corresponding layers are robust to identifier changes within each language.

Figure~\ref{fig:cross_lingual_similarity} illustrates cross-lingual semantic equivalence under syntactic variations, showing the average cosine similarity between representations of the same algorithm implemented in different languages. Again, both models reach their highest, or near-highest, similarity in intermediate layers: layers 8–15 for Llama-3.1-8B and a broader plateau across layers 31–43 for Qwen2.5-Coder-32B. While some curves peak slightly earlier or decay slowly, these ranges correspond to contiguous regions where semantic similarity is stably high across languages rather than to a single sharp maximum. This suggests that these layers encode language-agnostic, algorithm-level structure that abstracts away from surface lexical and syntactic choices.

Overall, the semantic probes indicate that the middle layers of both models develop language-agnostic representations that capture essential algorithmic logic while largely ignoring superficial variations.

\begin{figure}[t!]
    \centering
    \begin{subfigure}[b]{0.48\linewidth}
        \centering
        \includegraphics[width=\linewidth]{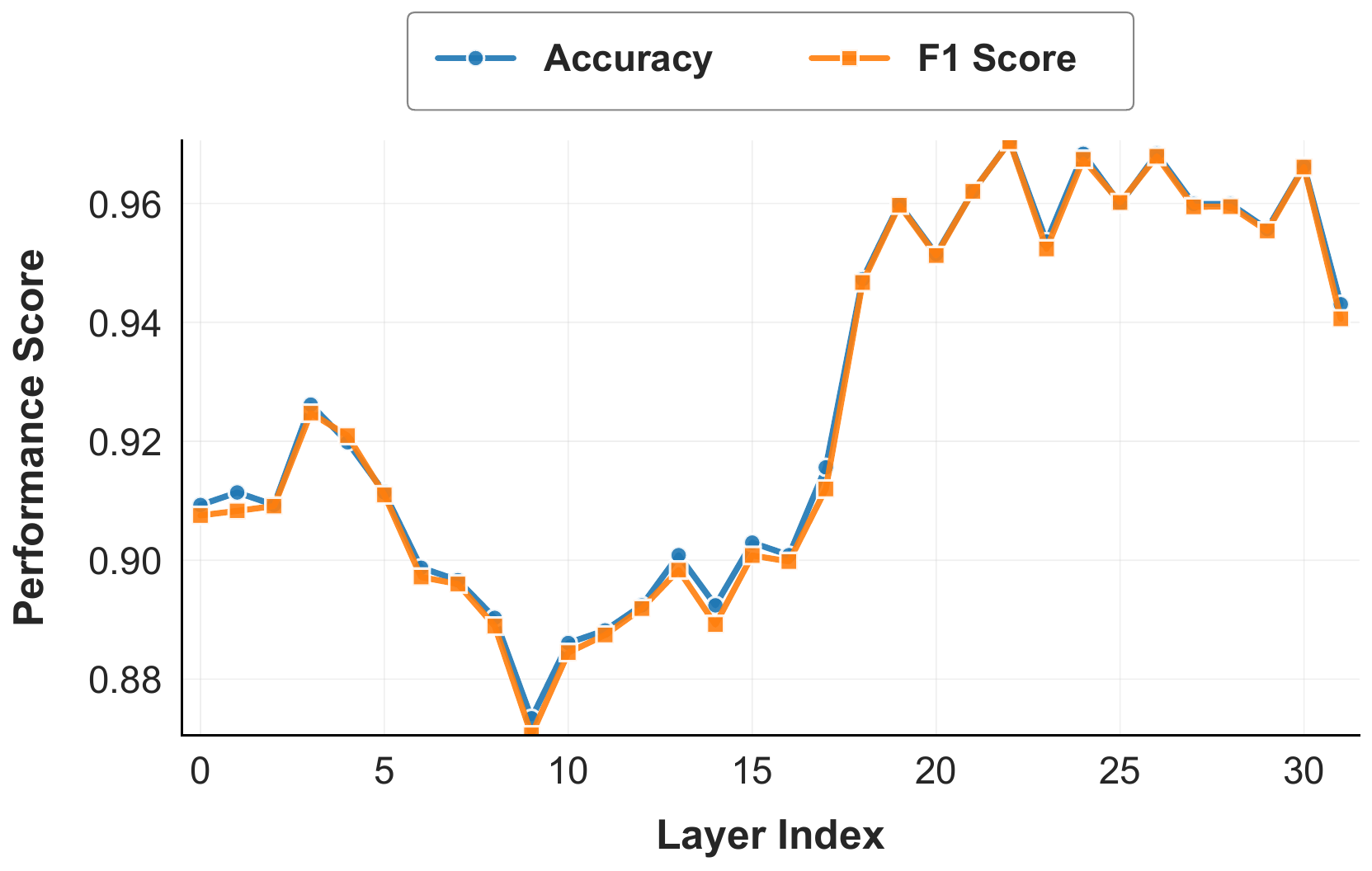}
        \caption{Llama-3.1-8B}
        \label{fig:ast_llama_separate}
    \end{subfigure}
    \hfill 
    \begin{subfigure}[b]{0.48\linewidth}
        \centering
        \includegraphics[width=\linewidth]{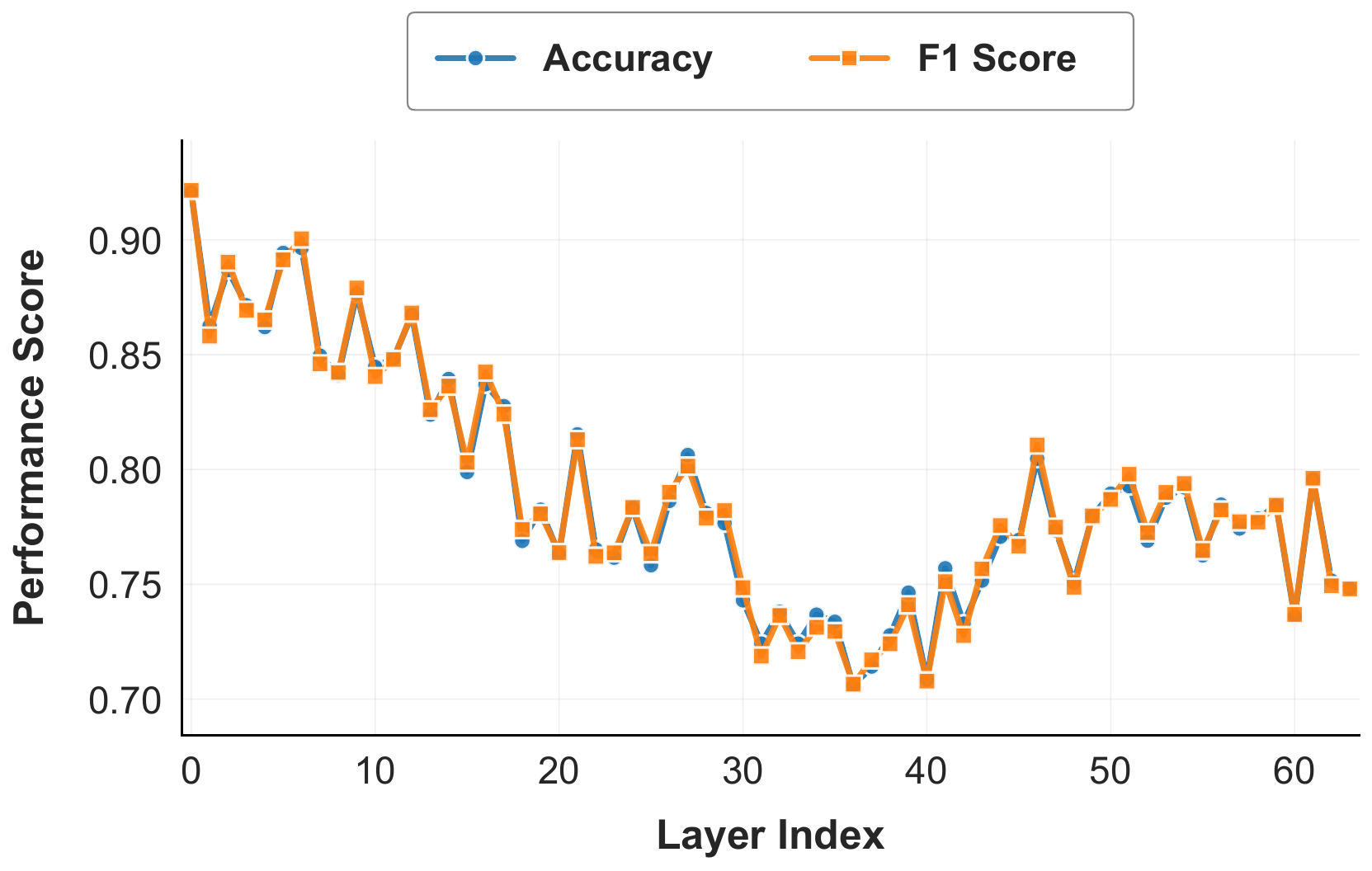}
        \caption{Qwen2.5-Coder-32B}
        \label{fig:ast_qwen_separate}
    \end{subfigure}
    \caption{
         Accuracy of AST Node Prediction across Model Layers. 
    }
    \label{fig:ast_probe_accuracy}
\end{figure}

\textbf{Results of Syntactic Structure Probing}. The decodability of explicit syntactic information follows a distinct U-shaped curve for both models, as shown in Figure~\ref{fig:ast_probe_accuracy}. The accuracy of AST node prediction is highest in the earliest layers, indicating that lower layers are strongly syntax-dominated and perform surface-level parsing. Performance then declines sharply, reaching its lowest point around layer 9 in Llama-3.1-8B and layer 36 in Qwen2.5-Coder-32B. This decline suggests that fine-grained syntactic features become less linearly separable and are encoded more abstractly in middle layers. Finally, accuracy recovers consistently in higher layers, consistent with the models' need to reconstruct syntactically valid output during code generation.

Taken together, Figures~\ref{fig:intra_lingual_similarity}–\ref{fig:ast_probe_accuracy} reveal an inverse relationship between semantic abstraction and syntactic fidelity across layers. Early layers combine relatively high semantic similarity with high AST prediction accuracy, reflecting strong surface-syntax encoding. In contrast, the middle layers are the first region where high semantic similarity coincides with minimal syntactic decodability, and we therefore designate these bands as the concept-layer regions of the models. Finally, upper layers recover syntactic information needed for generation, causing AST accuracy to rise again while semantic similarity gradually decreases.

More concretely, the convergence of peak (or near-peak) semantic similarity and minimum syntactic decodability delineates the concept layers. For example, in Llama-3.1-8B, the syntactic accuracy minimum at layer 9 falls within the semantic similarity band observed between layers 8 and 15. Similarly, in Qwen2.5-Coder-32B, the syntactic minimum at layer 36 lies inside the semantic plateau spanning layers 31 to 43. We therefore treat these intervals as approximate concept-layer regions where code is transformed into a universal, logic-centered representation before being converted back into language-specific syntactic structures in subsequent layers.

\begin{tcolorbox}[width=\linewidth, boxrule=0.8pt, left=2pt, right=2pt, top=2pt, bottom=2pt, colback=gray!5!white, colframe=gray!75!black]
\textbf{Finding 2:} Code language models organize knowledge hierarchically, forming distinct ``concept layers'' in their middle layers. Two complementary properties characterize these layers: they show peak invariance to lexical and syntactic variations, yet minimal retention of fine-grained syntactic details.
\end{tcolorbox}

\section{RQ3: Implications for Downstream Code-related Tasks} \label{sec:rq3}

To validate our findings and demonstrate the practical applicability of language-specific neurons (RQ1) and concept layers (RQ2) in software engineering, we design and conduct a series of targeted experiments across three code-related tasks: code generation, clone detection, and code summarization, as summarized in Table~\ref{tab:rq3_summary}.
\begin{itemize}
    \item\textbf{RQ3.1} \textit{Does a mechanistically-informed, neuron-guided fine-tuning strategy outperform LoRA for code generation in terms of both task performance and knowledge retention? (Section \ref{sec:RQ3.1})}
    
    \item\textbf{RQ3.2} \textit{Are the language-agnostic representations extracted directly from the concept layers effective for zero-shot code clone detection? (Section \ref{sec:RQ3.2})}
    
    \item\textbf{RQ3.3} \textit{Can the concept layers facilitate effective cross-lingual transfer learning for code summarization, improving performance on low-resource languages? (Section \ref{sec:RQ3.3})}
\end{itemize}

Our approach moves beyond treating the LLM as an opaque ``black box'', instead adopting a neuron-based mechanistic intervention strategy.

\begin{table}[t!]
\centering
\caption{Summary of Enhancement Techniques for RQ3 Tasks}
\label{tab:rq3_summary}
\begin{tabular}{@{}ll@{}}
\toprule
\textbf{Task} & \textbf{Enhancement Technique} \\ \midrule
Code Generation & Neuron-Guided Fine-Tuning \\
Code Clone Detection & Concept-Layer Embeddings (Zero-shot) \\
Code Summarization & Concept-Layer Guided Transfer Learning \\ \bottomrule
\end{tabular}
\end{table}


\subsection{Neuron-Guided Fine-Tuning for Code Generation}
\label{sec:RQ3.1}
\subsubsection{Study Design}
Parameter-efficient fine-tuning (PEFT) methods like LoRA are effective but typically modify a broad set of parameters without considering their functional roles. This may lead to suboptimal performance and catastrophic forgetting, where the model loses proficiency in previously learned tasks after fine-tuning~\cite{KirkpatrickPRVD16}. A more targeted, mechanistic fine-tuning approach can address these issues by preserving prior knowledge while improving target task performance.

In code generation, effective adaptation requires learning the syntactic structures and conventions of the target programming language. Based on our findings of language-specific neurons (RQ1), we propose a neuron-guided fine-tuning method that updates only the parameters in modules containing specialized neurons. This sparse fine-tuning~\cite{XuZMWC25} enhances performance on the target task while maintaining general code intelligence by preserving the rest of the model.

The procedure involves three steps:
\begin{enumerate}
\item \textit{Language-Specialized Neuron Identification}: Select neurons specialized for the target language using PLS scores;

\item \textit{Parameter Freezing}: Freeze all parameters except those in modules hosting the specialized neurons (\emph{e.g.}, linear layers like \texttt{gate\_proj});

\item \textit{Efficient Fine-Tuning}: Update only the unfrozen parameters using the MCEval instruction-tuning dataset.
\end{enumerate}

We compare our method with LoRA~\cite{HuSWALWWC22}, using Llama-3.1-8B as the base model due to resource constraints. To avoid potential contamination from benchmarks like HumanEval-X, we use the MCEval~\cite{ChaiL0YJLS0RGWW25} benchmark for both training and evaluation. Evaluation is based on two metrics: the \textit{tunable parameter ratio}, which reflects parameter efficiency, and the \textit{pass@3} metric, which measures the functional correctness of generated code. We also evaluate catastrophic forgetting by measuring the change ($\Delta$) in pass@3 scores, specifically assessing non-target languages (\emph{e.g.}, Python) after fine-tuning on a target language (\emph{e.g.}, Java). Superior forgetting mitigation is indicated by minimal degradation in non-target languages.

\settowidth{\myperfwidth}{\tiny\texttt{(-29.2\%)}}
\begin{table}[t]
\centering
\caption{Performance of Fine-Tuned Models for Code Generation (MCEval). The percentage change ($\Delta$\%) in pass@3 score relative to the base model is indicated with (\textcolor{teal}{↑} / \textcolor{red}{↓}).}
\label{tab:rq3_neuron_finetuning}
\small
\setlength{\tabcolsep}{3.5pt}
\newcommand{\perfchangecolored}[2]{%
    #1\% %
    \pgfmathsetmacro{\change}{(#1 - #2) / #2 * 100}%
    \def\myperfnum{\pgfmathprintnumber[fixed, zerofill, precision=1, showpos]{\change}\%}%
    \ifdim\change pt > 0pt%
        \def\myperfcontent{\textcolor{teal}{\texttt{(\myperfnum)}}}%
    \else\ifdim\change pt < 0pt%
        \def\myperfcontent{\textcolor{red}{\texttt{(\myperfnum)}}}%
    \else%
        \def\myperfcontent{\texttt{(\myperfnum)}}%
    \fi\fi%
    \makebox[\myperfwidth][r]{\tiny\myperfcontent}%
}
\resizebox{1\textwidth}{!}{
\begin{tabular}{llcccccc}
\toprule
\multirow{2}{*}{\textbf{Tuning Lang.}} & \multirow{2}{*}{\textbf{Tuning Method}} & 
\multirow{2}{*}{\makecell{\textbf{\% Tunable}\\ \textbf{Params}}} &
\multicolumn{5}{c}{\textbf{Pass@3 for Target Language}}  \\
\cmidrule(lr){4-8}
& & & \textbf{Python} & \textbf{Java} & \textbf{Go} & \textbf{C++} & \textbf{JS} \\
\midrule
\multicolumn{3}{l}{Base Model (without tuning)} & 36\% & 56\% & 48\% & 48\% & 38\% \\
\midrule
\multirow{2}{*}{Python} 
& LoRA & 0.53 &\cellcolor{gray!20}{\perfchangecolored{42}{36}} & \perfchangecolored{58}{56} & \perfchangecolored{34}{48} & \perfchangecolored{48}{48} & \perfchangecolored{34}{38}  \\
& Neuron-Guided & 0.48 &\cellcolor{gray!20}{\perfchangecolored{46}{36}} & \perfchangecolored{56}{56} & \perfchangecolored{38}{48} & \perfchangecolored{48}{48} & \perfchangecolored{36}{38}  \\
\midrule
\multirow{2}{*}{Java} 
& LoRA & 0.53 & \perfchangecolored{32}{36} & \cellcolor{gray!20}{\perfchangecolored{58}{56}} & \perfchangecolored{44}{48} & \perfchangecolored{48}{48} & \perfchangecolored{42}{38}  \\
& Neuron-Guided & 0.10 & \perfchangecolored{40}{36} & \cellcolor{gray!20}{\perfchangecolored{56}{56}} & \perfchangecolored{46}{48} & \perfchangecolored{46}{48} & \perfchangecolored{40}{38}  \\
\midrule
\multirow{2}{*}{Go} 
& LoRA & 0.53 & \perfchangecolored{44}{36} & \perfchangecolored{54}{56} & \cellcolor{gray!20}{\perfchangecolored{48}{48}} & \perfchangecolored{50}{48} & \perfchangecolored{34}{38}  \\
& Neuron-Guided & 0.03 & \perfchangecolored{42}{36} & \perfchangecolored{56}{56} & \cellcolor{gray!20}{\perfchangecolored{54}{48}} & \perfchangecolored{46}{48} & \perfchangecolored{36}{38}  \\
\midrule
\multirow{2}{*}{C++} 
& LoRA & 0.53 & \perfchangecolored{30}{36} & \perfchangecolored{52}{56} & \perfchangecolored{38}{48} & \cellcolor{gray!20}{\perfchangecolored{48}{48}} & \perfchangecolored{30}{38}  \\
& Neuron-Guided & 0.23 & \perfchangecolored{34}{36} & \perfchangecolored{54}{56} & \perfchangecolored{42}{48} & \cellcolor{gray!20}{\perfchangecolored{50}{48}} & \perfchangecolored{38}{38}  \\
\midrule
\multirow{2}{*}{JS} 
& LoRA & 0.53 & \perfchangecolored{32}{36} & \perfchangecolored{50}{56} & \perfchangecolored{42}{48} & \perfchangecolored{44}{48} & \cellcolor{gray!20}{\perfchangecolored{52}{38}}  \\
& Neuron-Guided & 0.12 & \perfchangecolored{38}{36} & \perfchangecolored{56}{56} & \perfchangecolored{40}{48} & \perfchangecolored{44}{48} & \cellcolor{gray!20}{\perfchangecolored{44}{38}}  \\
\bottomrule
\end{tabular}
}
\end{table}

\subsubsection{Results}
The code generation performance of our neuron-guided fine-tuning method compared to LoRA is shown in \Cref{tab:rq3_neuron_finetuning}. Our method achieves superior performance with a smaller fraction of tunable parameters. For example, when fine-tuned on Python, it achieves a 46\% \texttt{pass@3} score (+27.8\%), while LoRA achieves 42\% (+16.7\%). These results suggest that updating only the modules containing causally relevant neurons (as identified in RQ1) can match or exceed the performance of broader PEFT methods that update more parameters.

A key advantage of our method is its ability to mitigate catastrophic forgetting. While LoRA applies broad updates that can disrupt knowledge of non-target languages, our targeted approach better preserves this knowledge. On average, across five languages, our method shows only a 2.6\% decrease in performance on non-target languages, compared to 7.0\% with LoRA. The difference is most pronounced in C++: LoRA causes a 16.4\% degradation, whereas our method limits this to 5.4\%. These findings demonstrate that leveraging mechanistic insights about neurons offers a better trade-off between specialization and generalization.

\begin{tcolorbox}[width=\linewidth, boxrule=0.8pt, left=2pt, right=2pt, top=2pt, bottom=2pt, colback=gray!10, colframe=black]
\textbf{Finding 3.1:} Surgical fine-tuning of language-specific neurons outperforms standard LoRA with fewer updates while substantially mitigating catastrophic forgetting.
\end{tcolorbox}

\subsection{Clone Detection with Concept-Layer Embeddings}
\label{sec:RQ3.2}
\subsubsection{Study Design}
The detection of semantic (Type-IV) code clones, which are code fragments that share functionality despite substantial differences in implementation, syntax, and lexicon, remains a significant challenge~\cite{TaoZHX22}. Existing methods that rely on surface-level features often fail to capture this kind of deep semantic equivalence ~\cite{PinkuMR24}. The concept layers, which are identified in RQ2 as encoding abstract and language-agnostic representations of algorithmic logic, provide an ideal representational basis for this task.

The procedure consists of three steps:
\begin{enumerate}
    \item \textit{Representation Extraction}: For a given pair of code fragments, $(c_A, c_B)$, we process each using the frozen base model and extract the hidden-state vectors from the identified concept layers, $L_{\text{concept}}$.
    
    \item \textit{Embedding Aggregation}: We generate a single, fixed-size embedding vector for each fragment by applying mean pooling across the sequence of token vectors. This step yields the raw concept-layer embeddings:
    \begin{equation} \label{eq:concept_embedding}
    v_A = \text{Pool}(H_{L_{\text{concept}}}(c_A)) \quad \text{and} \quad v_B = \text{Pool}(H_{L_{\text{concept}}}(c_B)),
    \end{equation}
    where $H_{L_{\text{concept}}}(c)$ denotes the hidden states at the concept layers for code fragment $c$.
    
    \item \textit{Clone Classification}: Finally, we quantify the semantic similarity between the two fragments by computing the cosine similarity of their raw embeddings, $v_A$ and $v_B$. A pair is classified as a clone if this similarity score exceeds a threshold $\delta$, which is optimized on a designated validation set. 
\end{enumerate}

We assess the effectiveness of concept-layer embeddings for semantic code clone detection in a zero-shot setting using Llama-3.1-8B and Qwen2.5-Coder-32B as base models, with comparative analysis across other feed-forward layers. We also include an LLM prompting approach as an additional baseline, where the model is instructed to judge whether two code fragments are functionally equivalent. F1-score measures performance on the CodeNet\_clone~\cite{MoumoulaKKB25} benchmark. Strong zero-shot results indicate that the model acquires powerful and generalizable representations of code functionality~\cite{MoumoulaKKB25}, contrasting with fine-tuning-based methods~\cite{FengGTDFGS0LJZ20}.

\begin{figure}[t!]
    \centering
    \begin{subfigure}[b]{0.48\linewidth}
        \centering
        \includegraphics[width=\linewidth]{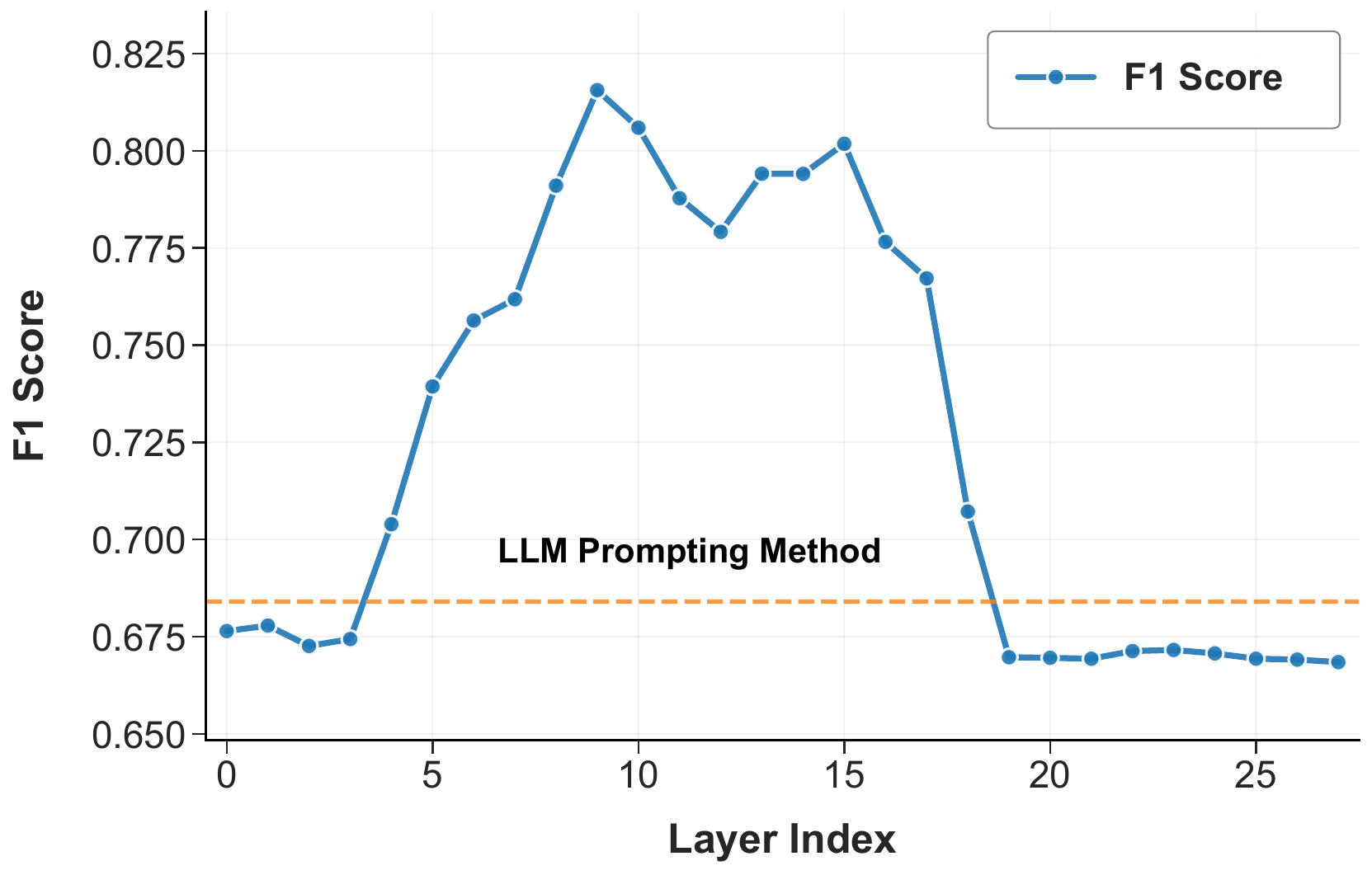}
        \caption{Llama-3.1-8B}
        \label{fig:ast_llama_separate}
    \end{subfigure}
    \hfill 
    \begin{subfigure}[b]{0.48\linewidth}
        \centering
        \includegraphics[width=\linewidth]{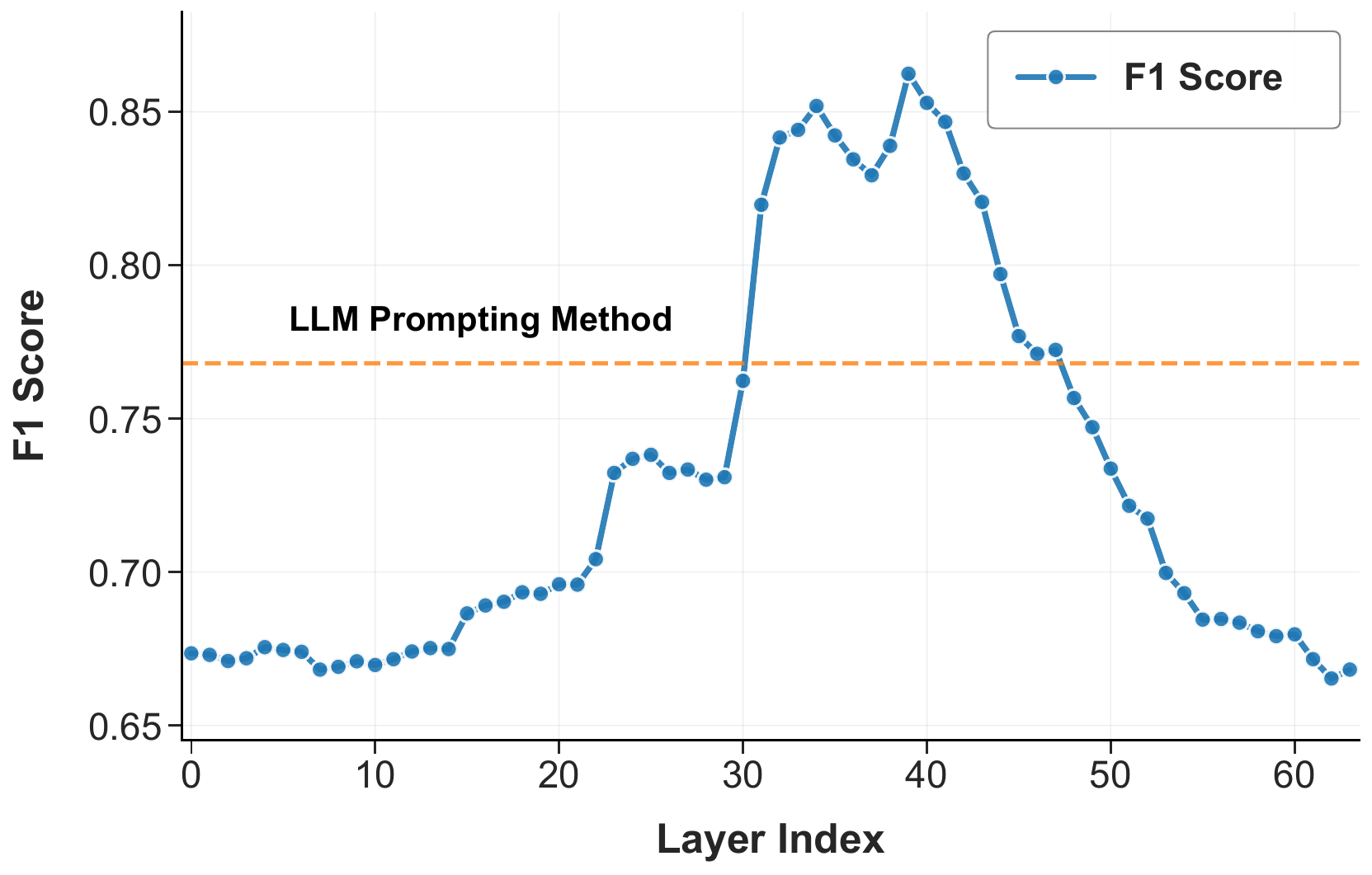}
        \caption{Qwen2.5-Coder-32B}
        \label{fig:ast_qwen_separate}
    \end{subfigure}
    \caption{F1-scores for Code Clone Detection using Embeddings across Model Layers.}
    \label{fig:clone_detection_f1}
\end{figure}

\subsubsection{Results}
\Cref{fig:clone_detection_f1} presents the F1-scores for zero-shot semantic code clone detection using embeddings from each layer of the language models. 
For Llama-3.1-8B, the performance peaks within its identified concept layers (layers 8--15), reaching a maximum of 0.823 at layer 9. This represents a 20.3\% relative improvement over the LLM prompting method of 0.684. 
A similar pattern is observed for Qwen2.5-Coder-32B, where performance also peaks within its concept layers (layers 31--43), achieving a maximum F1-score of approximately 0.86 at layer 38. This constitutes a 12.0\% relative improvement over its baseline of 0.768. 
The F1-score curves for both models illuminate a distinct functional hierarchy within their architectures: in the lower layers, performance is limited by lexical and syntactic representations, while the upper layers are more specialized for token generation and lack the abstraction required for clone detection. 

Notably, this strong performance is achieved in a zero-shot setting, without any task-specific fine-tuning. This indicates that the concept layers possess a powerful intrinsic generalization capability, mapping semantically equivalent code from different languages to proximal points in the representation space. Such inherent structure, optimized for abstract reasoning, makes the concept layers an ideal source of representations for semantic understanding tasks. Moreover, these findings provide strong evidence for our RQ2 hypothesis that the concept layers serve as the locus of language-agnostic logical encoding.

\begin{tcolorbox}[width=\linewidth, boxrule=0.8pt, left=2pt, right=2pt, top=2pt, bottom=2pt, colback=gray!10, colframe=black]
\textbf{Finding 3.2:} The concept layers act as a potent semantic hub. Their direct zero-shot representations demonstrate superior performance in semantic clone detection, significantly outperforming both holistic LLM prompting and embeddings from other layers.
\end{tcolorbox}

\subsection{Concept-Layer Guided Transfer Learning for Code Summarization}
\label{sec:RQ3.3}
\subsubsection{Study Design}
Generating high-quality natural language summaries for code in low-resource programming languages is challenging due to severe data scarcity and catastrophic forgetting of pre-trained knowledge~\cite{KirkpatrickPRVD16}. We leverage the concept layers (Section~\ref{sec:rq2}), which are language-agnostic representations of program logic, as a semantic bridge to enable cross-lingual transfer.

To adapt the model efficiently, we use LoRA~\cite{HuSWALWWC22} and insert trainable low-rank adapters into the self-attention modules of all concept layers and higher layers, while freezing all original parameters. This focuses parameter updates on high-level semantic processing and preserves syntactic knowledge acquired during pre-training.

Training proceeds in two stages:
\begin{enumerate}
    \item \textit{High-Resource Adaptation}: We first optimize the LoRA adapters on a large-scale code summarization dataset for high-resource languages (Python and Java), using cross-entropy loss over \texttt{docstring} tokens. This step teaches the model to map code semantics to natural language descriptions.
    \item \textit{Low-Resource Transfer}: We then continue training the same adapters on limited data from the target low-resource language (Ruby), again with cross-entropy loss. By reusing concept-layer representations learned on high-resource languages, the model can adapt to new syntax with substantially less data than training from scratch. This sequential approach offers a more efficient and direct form of cross-lingual transfer than conventional multilingual fine-tuning~\cite{TranTLG20, ConneauL19}.
\end{enumerate}

We compare our concept-layer guided transfer learning against four baselines:
(1) the Base model without any task-specific tuning;
(2) Direct, which performs single-stage LoRA fine-tuning on a broad set of layers independently for each language, using only that language's data;
(3) Concept-LoRA, which applies single-stage LoRA only to the identified concept layers and is trained solely on the target low-resource language (Ruby); and
(4) Two-Stage LoRA, a stronger baseline that follows the same high-resource$\rightarrow$low-resource schedule as our method but places LoRA adapters on broad layers instead of restricting them to concept layers.
As in previous experiments, Llama-3.1-8B serves as the base model, and we evaluate on CodeSearchNet~\cite{abs-1909-09436} using BLEU-4 and ROUGE-L.

\begin{table}[t]
\centering
\caption{Performance of Fine-Tuned Models for Code Summarization in Low-Resource Settings (CodeSearchNet)}
\label{tab:rq3_code_summarization}
\scriptsize
\setlength{\tabcolsep}{1.5pt}
\resizebox{\linewidth}{!}{%
\begin{tabular}{l|cc|cc|cc|cc|cc|cc}
\toprule
\multirow{2}{*}{\textbf{Method}} &
  \multicolumn{2}{c|}{\textbf{Python}} &
  \multicolumn{2}{c|}{\textbf{Ruby}} &
  \multicolumn{2}{c|}{\textbf{Go}} &
  \multicolumn{2}{c|}{\textbf{Java}} &
  \multicolumn{2}{c|}{\textbf{JS}} &
  \multicolumn{2}{c}{\textbf{PHP}} \\
\cmidrule{2-13}
 & {\tiny BLEU-4} & {\tiny ROUGE-L} & {\tiny BLEU-4} & {\tiny ROUGE-L} & {\tiny BLEU-4} & {\tiny ROUGE-L} & {\tiny BLEU-4} & {\tiny ROUGE-L} & {\tiny BLEU-4} & {\tiny ROUGE-L} & {\tiny BLEU-4} & {\tiny ROUGE-L} \\
\midrule
Base           & 11.00 & 24.84 & 9.80  & 21.56 & 12.57 & 26.19 & 10.61 & 24.02 & 9.76  & 21.08 & 12.21 & 27.41 \\
Direct         & 18.04 & 34.99 & 14.80 & 30.37 & 16.23 & 31.09 & 17.93 & 37.13 & 14.36 & 29.11 & \textbf{17.28} & \textbf{35.65} \\
Concept-LoRA   & 14.34 & 31.00 & 13.09 & 28.00 & 14.08 & 28.41 & 15.08 & 33.29 & 13.06 & 26.53 & 14.52 & 32.04 \\
Two-Stage LoRA & 17.88 & 34.87 & 14.66 & 30.18 & \textbf{19.08} & \textbf{40.27} & 17.19 & 35.73 & 14.05 & 27.62 & 15.46 & 33.03 \\
\textbf{Ours}  & \textbf{18.41} & \textbf{35.74} & \textbf{16.15} & \textbf{31.80} & 18.23 & 39.73 & \textbf{18.13} & \textbf{37.50} & \textbf{14.73} & \textbf{29.54} & 15.17 & 32.30 \\
\bottomrule
\end{tabular}
}
\end{table} 

\subsubsection{Results}
The code summarization results are reported in \Cref{tab:rq3_code_summarization}. For Ruby, our designated low-resource language, the proposed two-stage concept-layer transfer (Ours) achieves the best performance among all methods, with 16.15 BLEU-4 and 31.80 ROUGE-L. This corresponds to a 9.1\% improvement over Direct per-language LoRA (14.80 BLEU-4) and a 23.4\% relative improvement over the single-stage Concept-LoRA model trained only on Ruby (13.09 BLEU-4). It also surpasses the stronger Two-Stage LoRA baseline (14.66 BLEU-4, 30.18 ROUGE-L), indicating that focusing adaptation on concept layers yields additional gains beyond those provided by the high-resource$\rightarrow$low-resource schedule itself.

On high-resource Python, our method remains competitive or slightly better than the alternatives (18.41 / 35.74 BLEU-4 / ROUGE-L for Ours vs.\ 18.04 / 34.99 for Direct and 17.88 / 34.87 for Two-Stage LoRA). Across the remaining languages, Ours consistently outperforms the single-stage Concept-LoRA baseline and ranks among the top-performing methods, achieving the highest BLEU-4 in four out of six languages and at least the second-highest in one more. Overall, these patterns support the view that concept layers act as a semantic bridge: pre-adapting them on high-resource languages and then reusing the same adapters for a low-resource language yields more effective summarization in the low-resource setting, while maintaining competitive performance on high-resource languages.

\begin{tcolorbox}[width=\linewidth, boxrule=0.8pt, left=2pt, right=2pt, top=2pt, bottom=2pt, colback=gray!10, colframe=black]
\textbf{Finding 3.3:} Concept layers act as a semantic bridge for cross-lingual transfer: by first adapting them on high-resource languages and then reusing the same adapters for Ruby, our method substantially improves low-resource summarization compared with standard LoRA, direct in-domain tuning, and a stronger Two-Stage LoRA baseline.
\end{tcolorbox}


\section{Limitations and Threats to Validity} \label{sec:threats}

This study has several limitations that could affect the validity of our findings: 

\textbf{Polysemy in neurons.}
This study investigates the distribution of programming language-specific neurons in code LLMs by analyzing their activations under multilingual code inputs. However, prior work~\cite{Elhage2022Toy} indicates that individual neurons can be polysemous and may encode multiple distinct or unrelated concepts. Although the neurons identified here have been causally validated, their interpretation may remain incomplete and imprecise, since some neurons may participate in overlapping functionalities. A deeper examination of such polysemous neurons is reserved for future work.

\textbf{Generalizability of models and languages.} 
Our findings are derived from two specific dense transformer architectures and five programming languages. The extent to which these conclusions generalize to substantially different architectures (\emph{e.g.}, Mixture-of-Experts, models exceeding 70B parameters), proprietary closed-source models, or alternative programming paradigms warrants further investigation.
    
\textbf{Absence of ground truth.} 
Due to the lack of a standard ``ground truth'' for interpretability in code models, it is difficult to objectively evaluate explanation quality~\cite{ZhaoCYLDCWYD24}. To address this, we validate our results from two perspectives: (1) deactivating identified language-specific neurons and measuring their impact on code generation, and (2) demonstrating how the language-specific neurons and concept layer can inform training strategies and code presentation methods for multilingual code intelligence.

\section{Related Work}

Most interpretability efforts for code LLMs have centered on analyzing attention mechanisms \cite{SharmaCF022, abs-2207-08466, Saad024}. These studies examine attention weights and activations to understand where the model focuses throughout the input sequence. Mohammadkhani et al. \cite{MohammadkhaniTH23} analyzed attention scores of CodeBERT and GraphCodeBERT, revealing how these models allocate attention to different code parts in software engineering tasks. Liu et al. \cite{LiuTLL24} used attention interpretability to study how models like CodeT5 and CodeGPT attend to code elements during generation, translation, and repair. Paltenghi et al. \cite{PaltenghiP21} compared the attention patterns of neural code models to those of human developers. Wan et al. \cite{WanZZSXJ22} combined attention analysis, word embedding probing, and syntax tree induction to understand code structure and semantics.


Another strand of research focuses on representation and concept analysis to investigate how code LLMs internally organize and represent programming knowledge. Sharma et al. \cite{SharmaHQJ24} revealed how coding concepts are redundantly distributed across neurons. Ma et al.~\cite{MaLZ23} used Fisher information to demonstrate how multilingual code models group languages based on structural similarities. Haider et al. \cite{BlackBox} identified hierarchical concept encoding in feed-forward networks, ranging from syntax to semantics. Kargaran et al. \cite{Kargaran0YS25} used logit lens and neuron activation analysis on Llama models to study the shared representations between multiple programming languages and English. They found that the model's internal concept space is close to English and that while language-specific neurons are concentrated in the bottom layers, neurons exclusive to a single programming language tend to appear in the top layers. Wang et al.~\cite{WangWW23} extracted minimal critical code segments that drive model predictions. Troshin et al.~\cite{TroshinC22} quantitatively evaluated how program attributes are encoded across layers. 

Finally, causal and counterfactual analysis is emerging as an approach that seeks to identify cause-effect relationships in code LLM behavior through targeted interventions. Nader-Palacio et al.~\cite{NaderPalacioVCRMP24} established a formal basis for causal reasoning in code models. Gupta et al.~\cite{GuptaBJ25} extended this to multi-modal code generation, modeling how different inputs causally affect outputs. Cito et al.~\cite{CitoDM024} generated minimal code perturbations to reveal model decision boundaries. Hooda et al.~\cite{HoodaCAWFJ24} tested concept understanding by altering program logic and observing model responses. Rodriguez-Cardenas et al.~\cite{Rodriguez-Cardenas23} evaluated and compared causal methods for code model interpretation. 

Our work aligns firmly with the representation and concept analysis paradigm. In contrast to prior studies, we are the first to systematically analyze neuron activation patterns in response to multilingual code inputs and to assess the contributions of individual layers throughout the output generation process. 
Furthermore, we investigate the causal effects of identified programming language-specific neurons, and utilize these insights to strategically control model behavior, thereby enhancing performance on code-related tasks.

\section{Conclusion}
In this paper, we present a neuron-level interpretability study of code LLMs, uncovering the distinct roles of programming language-specific neurons and cross-lingual concept layers. Our findings offer deeper insights into the internal mechanisms underlying multilingual code understanding and generation. We further demonstrate the practical utility of these insights through neuron-guided fine-tuning, concept-aware embeddings, and transfer learning—consistently improving performance on tasks including code generation, clone detection, and summarization. By enhancing the transparency and controllability of code language models, this work provides a foundation for future research in explainable AI for software engineering and supports advances in multilingual code intelligence.

\section*{Data Availability}

All code and data used in this study are publicly available at: \url{https://github.com/mmuzhi/Neuron-Guided-Interpretation-of-CodeLLMs}. The repository includes: (i) the code for our proposed PLS method, the layer-wise probing framework (RSA and AST probes), and the downstream tasks; and (ii) all experimental datasets and results.

\begin{acks}
This research is funded by the National Key Research and Development Program of China (Grant No. 2023YFB4503802), the National Natural Science Foundation of China (Grant No. 62232003), and the Natural Science Foundation of Shanghai (Grant No. 25ZR1401175).
\end{acks}

\bibliographystyle{ACM-Reference-Format}
\bibliography{ref}

\end{document}